\documentclass[manuscript]{emulateapj}
\usepackage{natbib, hyperref}
\usepackage{amsmath}
\usepackage{color}
\usepackage{times}
\usepackage{graphicx}

\slugcomment{To appear in The Astronomical Journal}

\shorttitle{Young Star Cluster Complexes in NGC\,1427A}
\shortauthors{M.~D.~Mora et al.}

\begin{document}

\title{A Starburst in the Core of a Galaxy Cluster: The dwarf irregular NGC\,1427A in Fornax\footnotemark[1]\footnotemark[,2]}
\footnotetext[1]{Some of the data presented in this paper were obtained from the Mikulski Archive for Space Telescopes (MAST). STScI is operated by the Association of Universities for Research in Astronomy, Inc., under NASA contract NAS5-26555. Support for MAST for non-HST data is provided by the NASA Office of Space Science via grant NNX09AF08G and by other grants and contracts.}
\footnotetext[2]{Based on observations made with ESO Telescopes at the La Silla Paranal Observatory under programme ID 70.B-0695.}

%
 \author{Marcelo~D.~Mora\altaffilmark{3}}
\altaffiltext{3}{Instituto de Astrof\'isica, Pontificia Universidad Cat\'olica de Chile,
	Vicu\~na Mackenna 4860, 7820436 Macul, Santiago, Chile}

\author{Julio Chanam\'e\altaffilmark{3,4}}
\altaffiltext{4}{Millennium Institute of Astrophysics, Casilla 36-D Santiago, Chile}
\author{Thomas~H.~Puzia\altaffilmark{3}}

\begin{abstract}
  {Gas-rich galaxies in dense environments such as galaxy clusters and massive groups are affected by a number of possible types of interactions with the cluster environment, which make their evolution radically different than that of field galaxies. The dwarf irregular galaxy NGC 1427A, presently infalling toward the core of the Fornax galaxy cluster for the first time, offers a unique opportunity to study those processes in a level of detail not possible to achieve for galaxies at higher redshifts, when galaxyÐscale interactions were more common. Using the spatial resolution of Hubble Space Telescope/Advanced Camera for Surveys and auxiliary Very Large Telescope/FORS1 ground-based observations, we study the properties of the most recent episodes of star formation in this gas-rich galaxy, the only one of its type near the core of the Fornax cluster. We study the structural and photometric properties of young star cluster complexes in NGC 1427A, identifying 12 bright such complexes with exceptionally blue colors. The comparison of our broadband near-UV/optical photometry with simple stellar population models yields ages below $\sim4 \times 10^{6}$ years and stellar masses from a few 1000 up to $\sim 3 \times 10^{4} M_{\odot}$, slightly dependent on the assumption of cluster metallicity and initial mass function. Their grouping is consistent with hierarchical and fractal star cluster formation. We use deep H$\alpha$ imaging data to determine the current star formation rate in NGC 1427A and estimate the ratio, $\Gamma$, of star formation occurring in these star cluster complexes to that in the entire galaxy. We find $\Gamma$ 
  {{to be among the largest such values available in the literature, consistent with starburst galaxies.  Thus a large fraction of the current star formation in NGC 1427A is occurring in star clusters, with the peculiar spatial arrangement of such complexes strongly hinting at the possibility that the starburst is being triggered by the passage of the galaxy through the cluster environment.
  }}} 

\end{abstract}



\section{Introduction}
Star clusters are among the most common stellar systems in the universe, present in almost all galaxies and systematically studied during the last decade thanks to the spatial resolution of the {\it{Hubble Space Telescope}} \citep[HST;][]{seth04, lotz04, sharina05,hasegan05, jordan05, jordan06, jordan07, mieske06, peng06, peng08,
sivakoff07, georgiev08, georgiev09b, georgiev09a, georgiev10,masters10, villegas10, liu11, wang13}.  Their formation occurs as part of a hierarchical process where large interstellar regions fragment into giant molecular clouds and cloud cores  \citep[e.g.][]{Elmegreen:2011fk_1}. These fragments can be highly sub-structured, containing dense clumps and filaments showing a typical fractal dimension \citep[e.g.][]{Scheepmaker:2009pd, Sanchez:2008kx} where clusterÐcluster interactions occur and may lead to star-cluster disruption or star-cluster merging  \citep{Kroupa1998,Fellhauer2005}.~In large numbers star clusters have been detected in violent environments \citep[e.g.][etc]{Whitmore2005,Scheepmaker:2009pd,Chandar:2011zr,Mullan:2011lq}  and in small numbers in isolated environments, like late-type galaxies  
\citep[e.g.][]{Eskridge:2008ty} and dwarf galaxies \citep[e.g][]{Billett:2002yu} showing that the environment plays a significant role in the intensity of their formation. 
\citep[See][for a typical formation rate in isolated environments for dwarf galaxies]{Cook:2012yq}. Observational evidence has shown a clustering of young star clusters, originally described as Òknots,Ó within disks in the Antennae\citep{Whitmore:1999fk}, which are formally referred to as star cluster complexes
  \citep[e.g.][]{Adamo:2012nx}.  The location of the star cluster complexes seems to be related to spiral arms and the tidal streams in between interacting galaxies, as has been observed, e.g., in M51, where a strong correlation with H$\alpha$ and 20 cm radio-continuum emission, decreasing for older clusters, has also been reported \cite{Scheepmaker:2009pd}. 
  Similar results were obtained in the Milky Way: open clusters tend to cluster in groups when age, spatial distribution, and kinematics are taken into account simultaneously \citep{de-la-Fuente-Marcos:2008ve} suggesting a common past in a larger hierarchical structure. The further evolution of star cluster complexes will make them disappear on certain timescales as it is the case in the LMC \citep{Gieles:2008ve}. Similar behavior, was observed in several dwarf galaxies by \cite{Bastian:2011rr}.  In general, the clumping of structures seems to be universal where the star cluster complexes themselves play a role in triggering star formation in isolated locations, like the end of tidal stream, as is seen in the Ruby Ring, where the central cluster triggered the formation of a ring of clusters \citep{Adamo:2012nx}.%

~In this paper, we report the characterization of young star cluster complexes in NGC1427A, a dwarf irregular (dIrr)
galaxy infalling toward the center of the Fornax galaxy cluster at a mildly supersonic speed \citep{Chaname:2000ve}. 
One of the earliest ground-based photometric studies of this galaxy was performed by  \cite{Cellone1997}, who already observed that the galaxy has features similar to interacting galaxies, such as a distorted ring-like structure mainly composed of blue young objects. They also noted that a bright star cluster complex in the northern part of the galaxy may be a distinct Fornax object in the vicinity of NGC 1427A, as a consequence of a past interaction.  \cite{Chaname:2000ve} 
showed, based on long-slit spectroscopic observations, that the northern object has the same velocity pattern as the rest of the galaxy, discarding the hypothesis of a separate entity.

Based on multi-passband ground-based imaging data,  \cite{hilker1997} 
pointed out that a majority of bright OB associations and HII regions in NGC 1427A were aligned along its southÐwest edge, and reported the detection of about 34 cluster-type objects uniformly distributed over the entire galaxy. Finally, using near-UV to near-IR VLT imaging, \cite{Georgiev2006} reported the
detection of $38\pm8$ globular clusters candidates with colors
compatible with an old metal-poor population ($\sim\!10$ Gyr and
$0.4~Z_{\sun}\!<\!Z\!<\!Z_{\sun}$) and a specific frequency $S_N\!=\!1.6$, placing NGC 1427A among the mean LMC-type galaxies in
terms of their old star cluster content \citep{georgiev10}.

The above previous works all hypothesize that NGC 1427A is experiencing enhanced star formation and morphological distortions induced by its interaction with the galaxy cluster environment. Although the details of such interaction need to be elucidated, it most likely leads to the cessation of star formation within the galaxy before a few cluster crossing times. Such events seem to be more common at higher redshifts, as demonstrated by \citet{Mahajan2012} in their study of star cluster formation in the outskirts of galaxy clusters. Given the proximity of the Fornax cluster, NGC 1427A thus provides a unique opportunity to study the last star cluster formation episode(s) in a gas-rich galaxy before the residual star- formation fuel is removed and star-cluster formation is altogether quenched.

Our paper is structured as follows. In Section 2, we describe the data, the reduction procedure, photometry and we test the limit of our sample through completeness tests. In Section 3, we describe our results, and in Section 4 we discuss our findings and draw conclusions.


\section{OBSERVATIONS AND DATA REDUCTIONS}

The present work is based on a combination of space- and ground-based observations. Space-based observation of NGC 1427A were acquired using the Advanced Camera for Surveys (ACS) on board of the $HST$ (PI: Gregg, proposal ID: 9689) through the filters F475W, F660N, F625W, F775W, and F850LP. For each pointing, three exposures with a small offset were acquired. A journal of all the $HST$ observations is given in Table 1. Ground-based observations were taken in service mode using the FORS1 instrument mounted at the Very Large Telescope (VLT) on Cerro Paranal in Chile (PI: Reisenegger, proposal ID: 70.B-0695). From this observing run we only use the U-band images. More details of the VLT U-band observations in table N\,1 of \cite{Georgiev2006}.

\begin{figure}
\centering
\includegraphics[angle=0,scale=0.25]{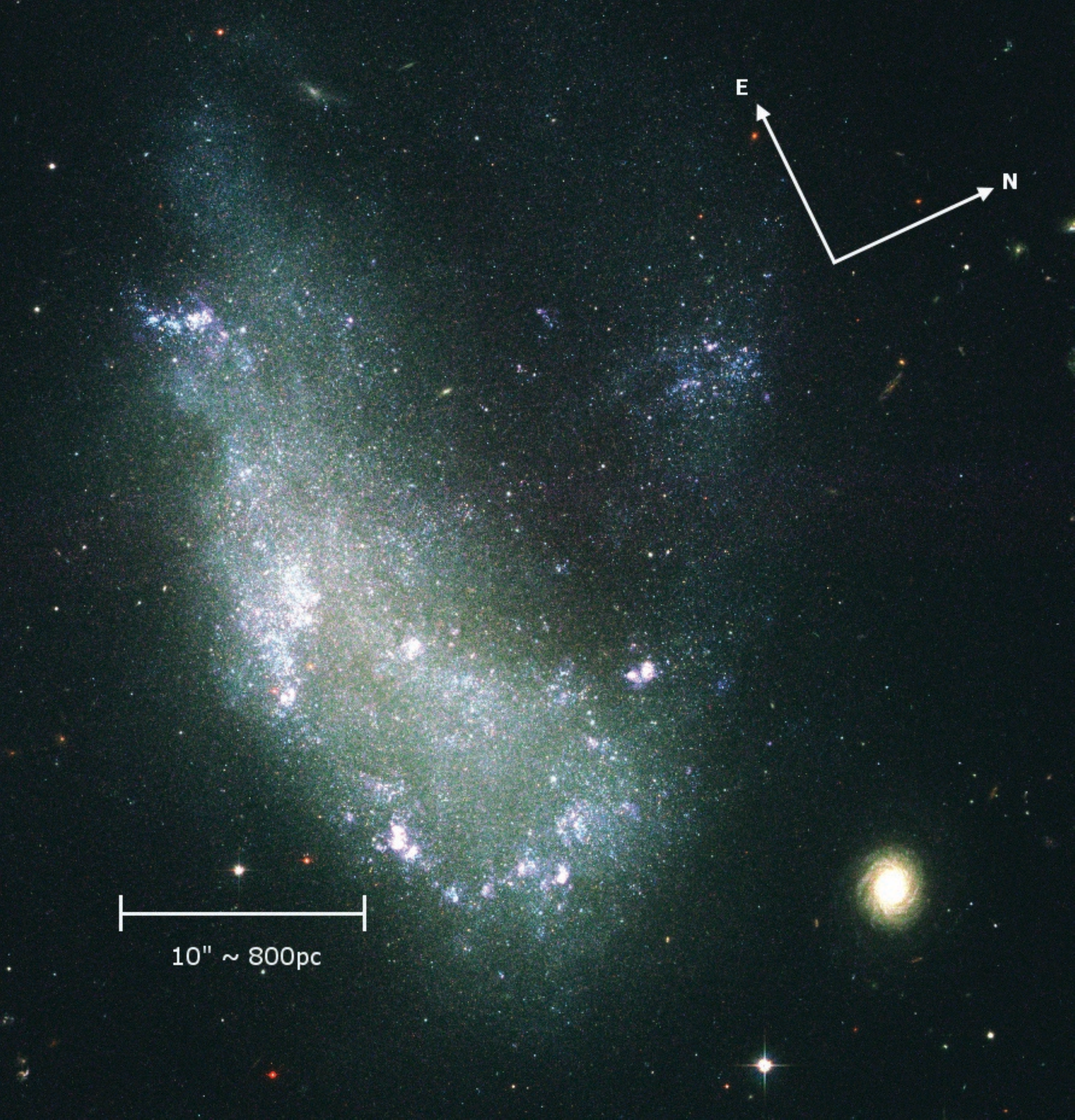}
\caption{
Color composite HST/ACS image of NGC~1427A. For the blue, we used the F475W channel; for the green, we used the F625W and F775W channels; and for the  red we used the F850LP  and the F660N channels. The image correspond to a field of view of $\sim$124$\arcsec \times$~145$\arcsec$. The arrow indicates the north and east. 
}
\label{FigNGC1427A}
\end{figure}

\begin{deluxetable}{lcccc}[!ht]
\tabletypesize{\scriptsize}
\tablecaption{Journal of HST observations. \label{TABLA_EXPOSURE}}
\tablewidth{0pt}
\tablehead{
\colhead{Filter} &  \colhead{Observing Date} &\colhead{Camera} &  \colhead{$t_{\rm exp}$ [sec]} & \colhead{Zero Point}
}
\startdata
             F475W &   2003 Jan 9 & ACS/WFC &  1960 &26.176\\
             F625W &   2003 Jan 9 & ACS/WFC &  1820 &25.751\\
             F660N &    2003 Jan 9 & ACS/WFC &  1790 &22.114$^{a}$\\
             F775W &   2003 Jan 9 & ACS/WFC &  1430 &25.283\\
             F850LP &  2003 Jan 9 & ACS/WFC &  1400 &24.351
\enddata
\tablecomments{$^{a}$STmag.}
\end{deluxetable}

\subsection{ACS data: Object Detections}
The $HST/ACS$ images were downloaded from the $HST$ Archive in the pre-processed {\sf \_flt} format.~The images were charge transfer efficiency corrected and each filter was combined into a single final image with the {\sc MultiDrizzle} routine \citep{koekemoer} using a lanczos3 kernel function with the sky subtraction option disabled. The final color composite image is shown in Fig \ref{FigNGC1427A}. 

At the distance of NGC\,1427A \citep[$D\!=\!19\!\pm\!1.8$\,Mpc][]{HST_KEY_PROJECT}, one ACS/WFC pixel corresponds to $\sim $4 pc and one pixel in the FORS images corresponds to $\sim$16 pc.~Therefore, due to the superior $HST$ resolution, we use the ACS/WFC frames for object detection.~However, since we do not know the optimal number of connected pixels that maximizes object detections, 
we performed several detection tests using {\sc SExtractor} ({v2.8.6})~software package \citep{SExtractor} aiming at maximizing the number of marginally resolved and unresolved objects in all filters. We took  the photometrically deepest frame, F475W, and convolved the image with synthetic gaussian PSFs of various FWHMs. We then ran {\sc SExtractor}, recording each time the number of detections.~After several runs, we found that the largest number of point-sources and slightly extended objects was detected in the F475W image convolved with a point-spread function (PSF) of FWHM=2.5 pixels. Therefore, based on this experiment, we considered an area of three connected pixels with 4$\sigma$ above the background noise level as our object detection setup in the original, i.e. non-convolved, F475W frame.

\subsection{ACS data: Photometry and Sizes}
\label{ln:photsize}

To account for the variable PSF across the ACS field, we used the effective point-spread function ({ePSF}) technique \citep{EPSF:EPSF}. Briefly, the ePSF technique uses fiducial PSFs distributed across each WFC chip \citep[see Fig 2;][]{EPSF:EPSF} which were added to empty {\sf \_flt} images and later drizzled into a final single image repeating the exact procedure  it was done for the science images using the {\sc Multiking} ACS/$HST$ simulator \citep{Paolillo:2011fk}.~Therefore, our ePSFs are affected by the same distortion corrections as the science images \citep[for a detailed discussion of the ACS/WFC resolution limitations see][]{Puzia:2014qy} .~Typical variation for the contiguous PSF FWHM were $\leq$ 0.2 pixels and largest difference was observed between each corner of the ACS at $\sim 0.6$ pixels.~Because we are aiming at measuring sizes and magnitudes, each ePSF was extracted from the drizzled image using the {\sc Iraf}\footnote{IRAF is distributed by the National Optical Astronomical Observatory, which is operated by the Association of Universities for Research in Astronomy, Inc, under cooperative agreement with the National Science Foundation.} {\sc PSF} task and later subsampled by a factor of 10 with the {\sc seepsf} {\sc Iraf} task.~Object sizes and magnitudes were measured using {\sc Ishape} \citep{Ishape}.~Briefly, {\sc Ishape} utilizes an analytic function convolved with a PSF to model  
each source in an iterative routine in which the  shape parameters, like the FWHM, are adjusted until the best fit is obtained.~  In our case, we used as analytic function  a King profile \citep{King}  considering a concentration parameter $c$, which is the ratio between tidal and core radius, fixed to $c=30$ which was later convolved with the nearest ePSF position. We chose this profile in order to make this work compatible with previous studies.
From {\sc Ishape} we obtained object sizes, fluxes and signal-to-noise ratio (S/N) values. Since S/N correlates with the sky  background,  we  have calculated the magnitude errors  following the formula:  $2.5\times \log{_{10}{(1+\mathrm{N}/\mathrm{S}})}$.~ACS zeropoints were adopted from the $HST$/ACS web pages using the "WFC and HRC Zeropoints Calculator"\footnote{{http:{\slash}{\slash}www.stsci.edu{\slash}hst{\slash}acs{\slash}analysis{\slash}zeropoints{\slash}zpt.py}} considering the date of the image observation.

\begin{figure}
\centering
\includegraphics[width=9cm]{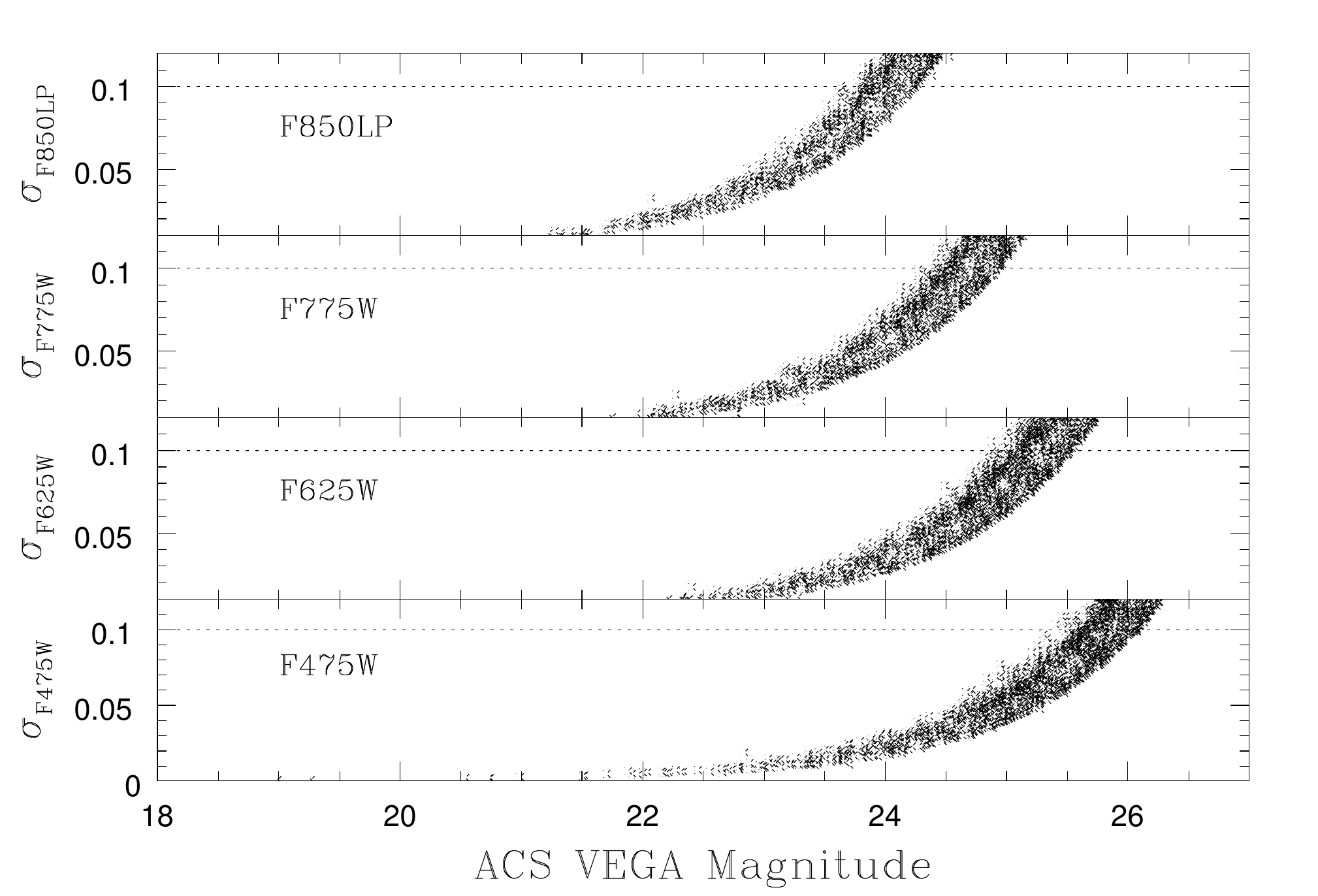}
\caption{HST/ACS Vega magnitudes vs magnitude errors for four ACS filters. Dashed lines correspond to error limits of $\sigma_{\mathrm{ACS-Filter}} \le 0.1 mag$. %
}
\label{MAG_ERR}
\end{figure}
    
In Figure~\ref{MAG_ERR}, we present the measured ACS magnitudes versus photometric errors of detected objects in four filters. The F475W photometry is the deepest compared with the other three ACS filters, justifying this filter as our detection frame. To base our analysis on the highest quality photometry, we will consider in the following only those objects with an error in magnitude less than 0.1 mag in all four ACS filter.

\begin{figure}
\centering
\includegraphics[width=9cm]{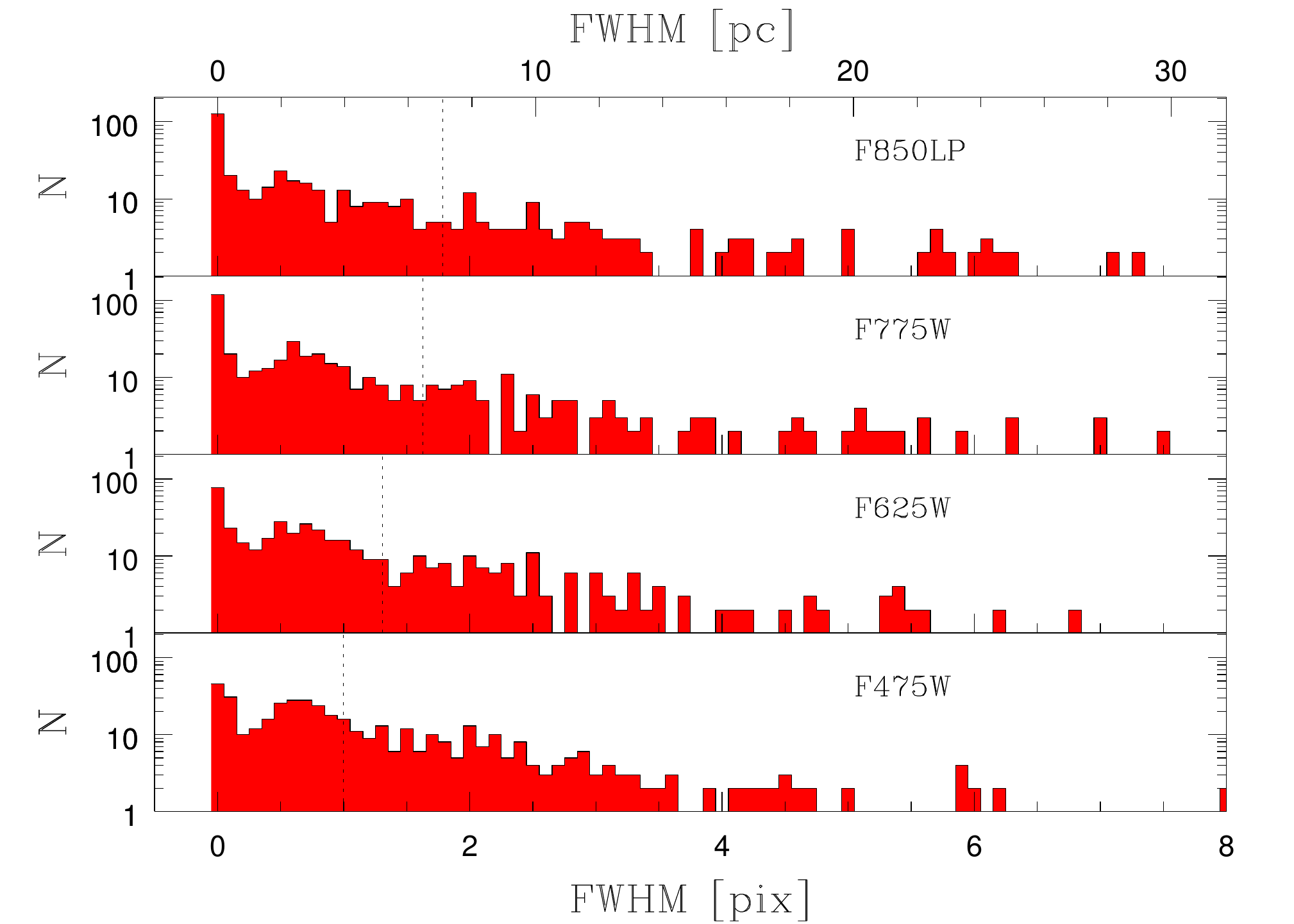}
\caption{Object sizes measured by {\sc Ishape}. Solid histograms correspond to the FWHM of objects with an $\sigma_{mag} \le 0.1$ mag. Peaks at zero correspond to unresolved objects. Bin size corresponds to 0.1~pix.~Dashed lines correspond to the nominal diffraction limit ($1.22 \lambda/D$): F475W: 3.95 pc, F625W: 5.20 pc, F775W: 6.45 pc, and F850LP: 7.07 pc. }
\label{FWHM}
\end{figure}

Because we use {\sc Ishape} to measure star cluster sizes, we can study their size distributions in each filter. Figure~\ref{FWHM} illustrates the distribution of FWHM values (in pixels/parsecs).~Since {\sc Ishape} takes the PSF directly into account, the histogram shows peaks at FWHM~$\!=\!0$ pc, which correspond to unresolved, point-source-like objects. Due to the larger diffraction limits of redder filters, a larger number of point-like objects is expected in the reddest filter compared with the bluest filter, fully consistent with our observations.

\subsection{$U$-based FORS1 photometry}

\begin{figure}
\centering
\includegraphics[width=8.5cm]{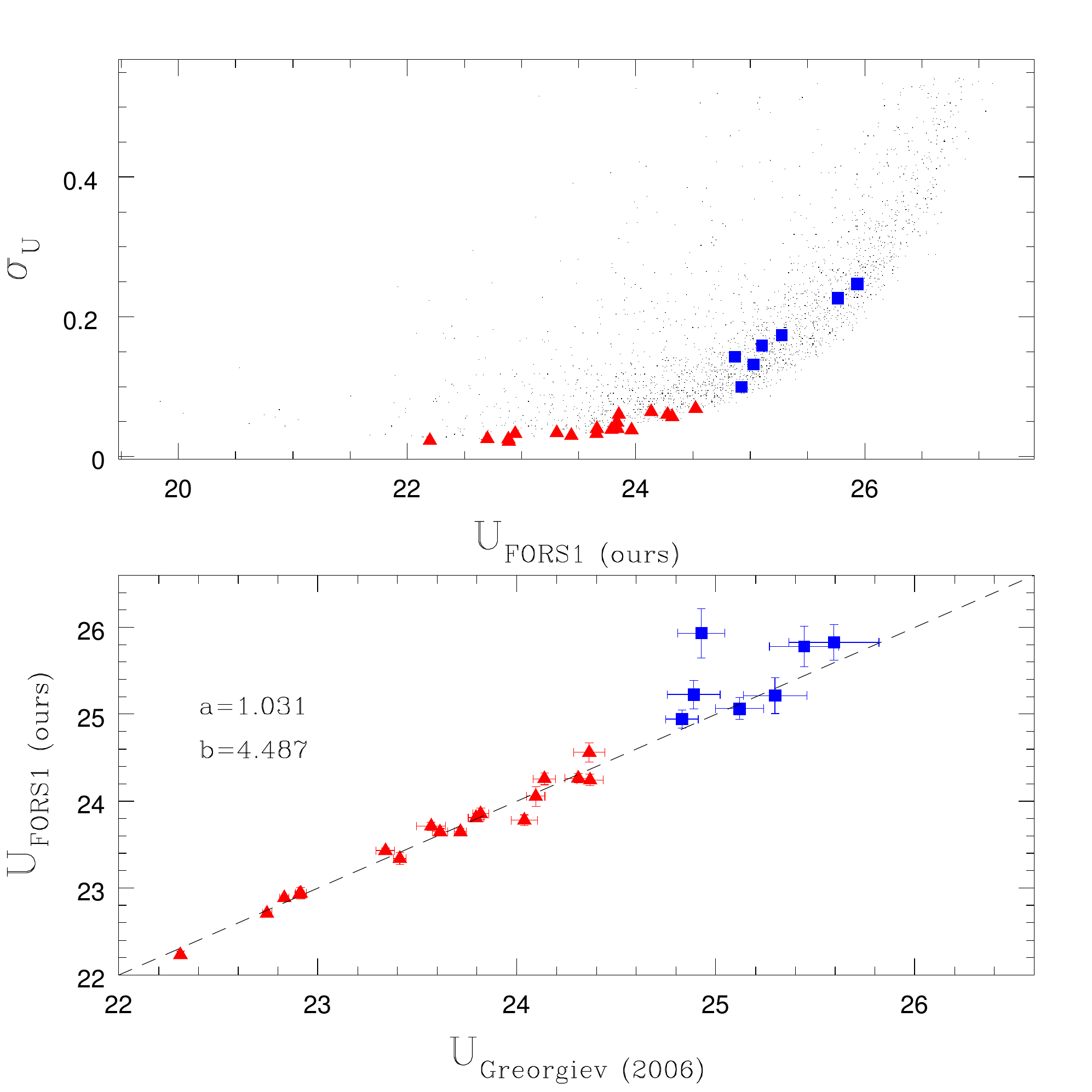}
\caption{(Top panel): FORS1 U-band photometric errors as a function of calibrated U-band magnitude using \cite{Georgiev2006} as reference. Triangles and squares correspond to the same objects as the bottom panel. (Bottom panel): photometry comparison between our work and \cite{Georgiev2006}.~Triangles and squares indicate matched globular clusters between the two studies. Because of the smaller photometric errors (i.e. $\sigma_{\rm U}\!<\!0.1$ mag), we only used the brighter objects (red triangles) for the derivation of the photometry fit, and only for completeness purposes we show the objects with larger photometric errors (blue squares). The dashed line corresponds to the one-to-one relation and is plotted for guiding purposes.}
\label{Ubandphot}
\end{figure}

U-band FORS1 images were bias and flat-field corrected and later combined into a deep master U-band image. Since the object detection was performed in the ACS frame, we transformed the $HST$/ACS coordinates into the master U  frame using the {\sc Geomap} and {\sc Geoxytran} tasks in the {\sc Iraf} environment. Since {\sc Ishape} fits the central position of the objects, and due to the lower spatial resolution of FORS1, we used PSF photometry because we can deactivate the center fit option and thus use directly the positions from the ACS-FORS1 coordinate transformations.
We then conduct a preliminary photometry run using the {\sc Phot} task in {\sc Iraf}. Ten bright and isolated sources were used to construct an empirical PSF with 20 pixel radius using the {\sc Psf} task. This empirical PSF is later used to carry out PSF photometry with the {\sc Allstar} task. Finally, the photometric calibration of the U-band images was performed by means of the comparison between the calibrated photometry of the globular clusters from \cite{Georgiev2006} and the uncalibrated matched globular clusters detected by us in our master U-band frame. The reason to select them for comparison is because they are isolated objects. Therefore we minimize possible nearby object contaminations.  In Figure~\ref{Ubandphot} we show the comparison between our
photometry and that of  \cite{Georgiev2006}, as well as the U-band photometric error as a function of magnitude.~Matched globular clusters are shown with filled triangles and filled squares. However, since faint globular clusters show larger errors than the bright ones we decided to calibrate the photometry using only globular clusters with an error less than 0.1 mag.~These objects are plotted as filled red triangles in Figure~\ref{Ubandphot}.~A function of the form $y\!=\!ax\!+\!b$ was fitted to the bright globular clusters for the photometric calibration. We obtain the following values  
{$a\!=\!1.03\!\pm\!0.03$ and $b\!=\!4.5\!\pm\!0.4$}. 
For completeness purposes, we are plotting the faint globular clusters that were not used for the U-band calibration as blue squares. Despite the fact that we did not use the faint globular clusters for the photometric calibration, they follow the one to one line comparison and are in agreement with the photometric calibration.

\subsection{Previous Ground-based Photometry versus $ACS$ Photometry}
\begin{figure*}
\centering
\includegraphics[width=5.9cm]{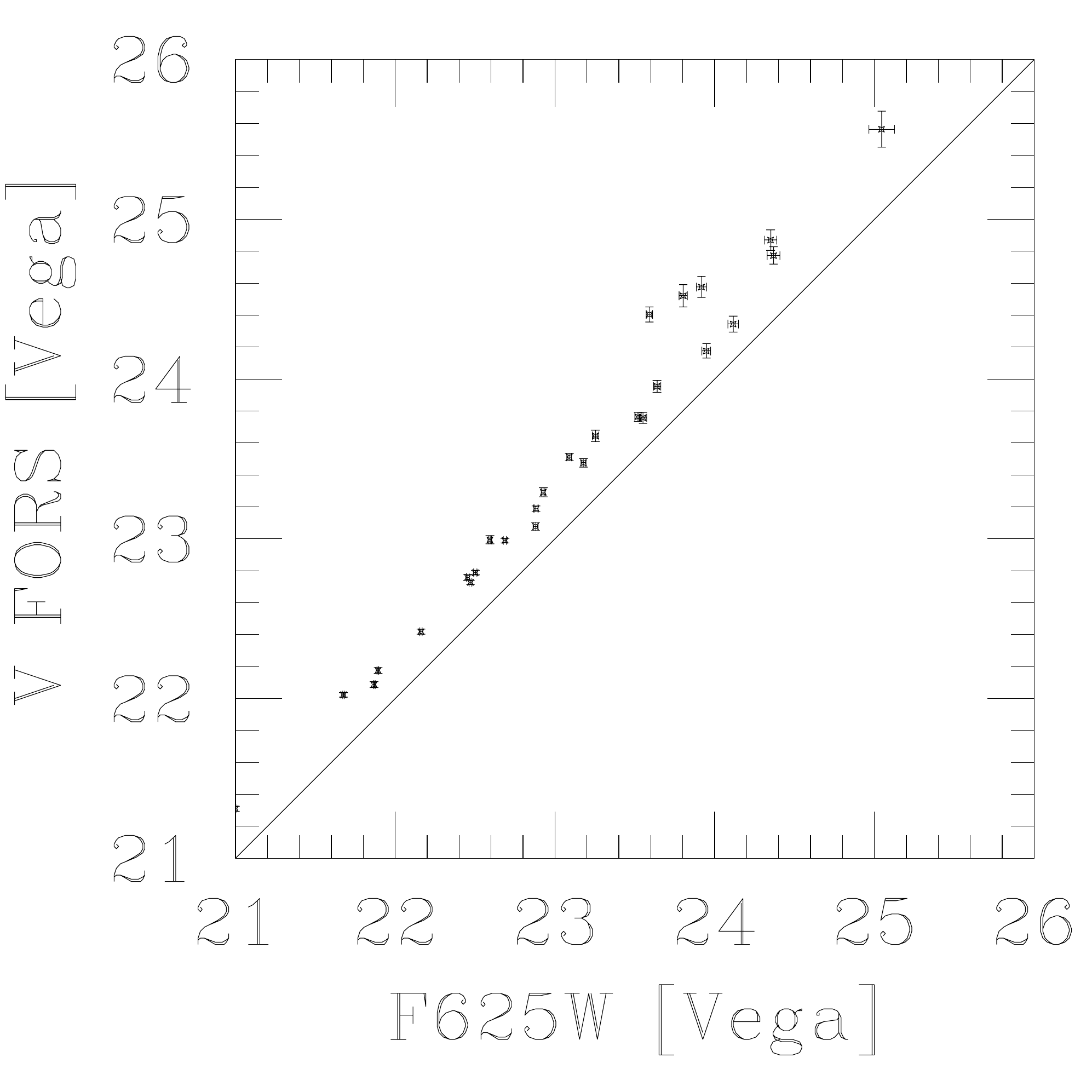}
\includegraphics[width=5.9cm]{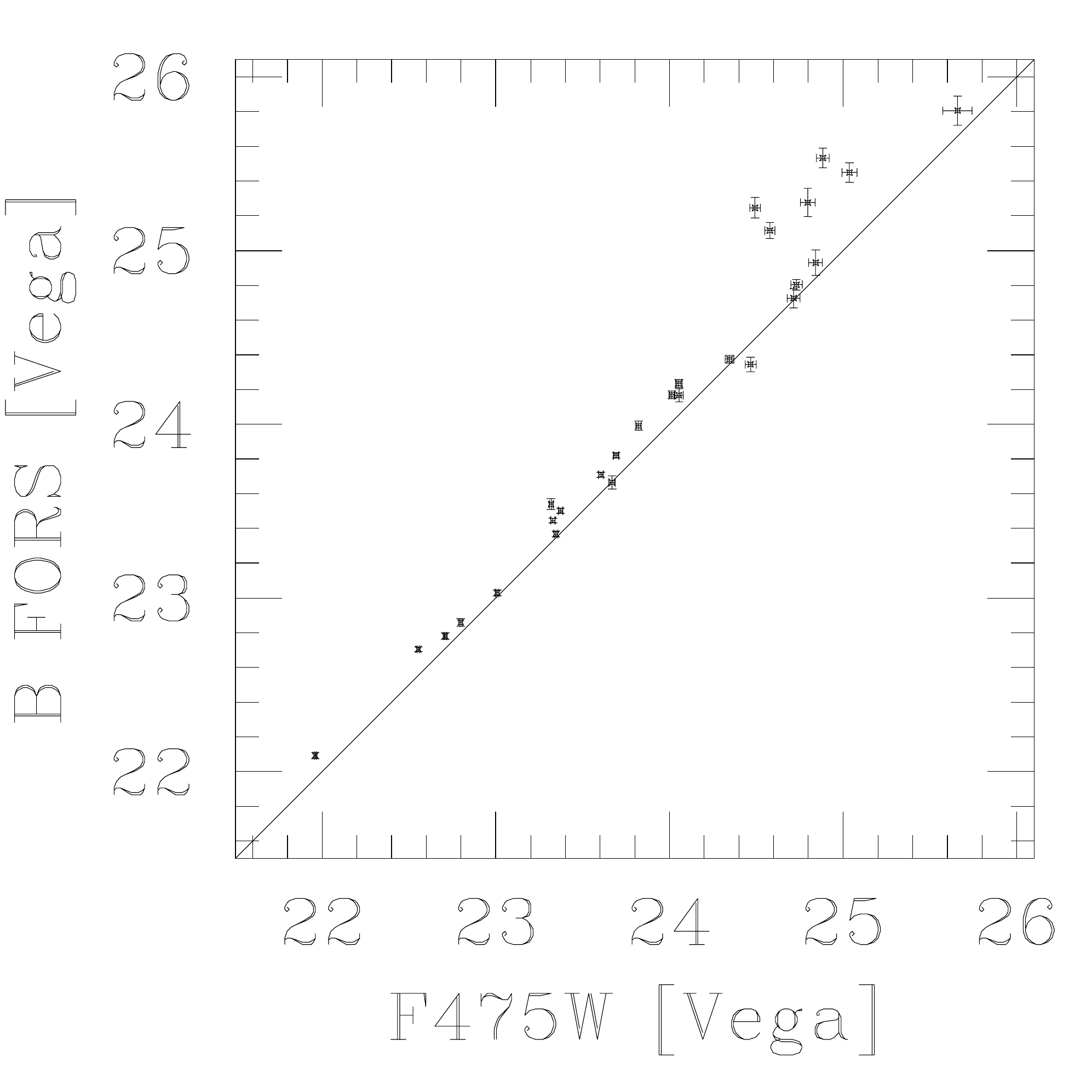}
\includegraphics[width=5.9cm]{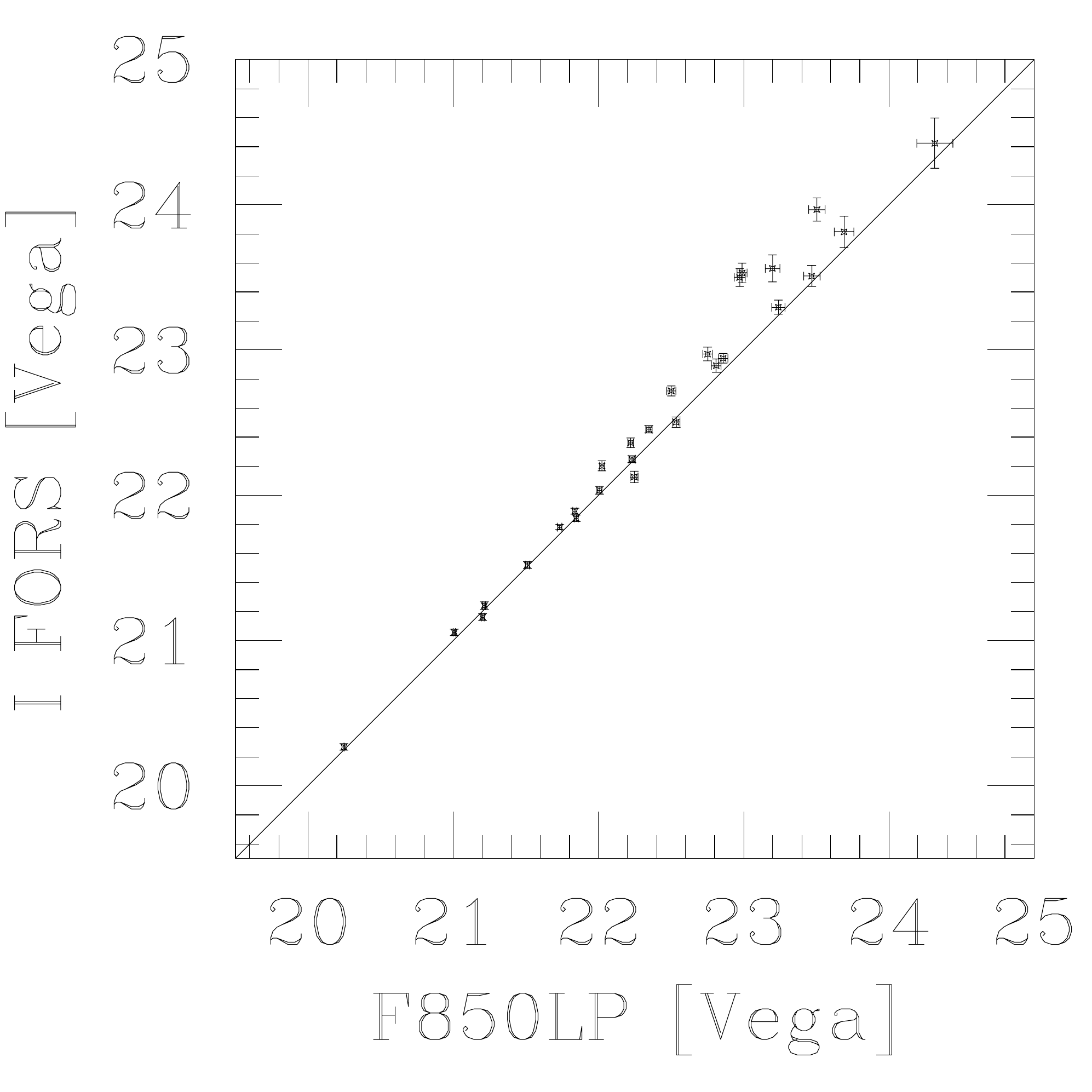}

\caption{Photometry comparison between our work and that of \cite{Georgiev2006}. (Left panel): V-band against F625W. (Central panel): B-band against F475W. ( Right panel): I-band against F850LP photometry. } 
\label{ACS_FORSbandphot}
\end{figure*}

In order to double check our ACS photometry, we compared the globular clusters detected on the ACS images with its counterparts from \cite{Georgiev2006} measured by FORS1. Since the ACS passbands do not have the same wavelength coverage nor the same shape as the FORS1 filters, the comparison is not exact, and would require a color term. Figure~\ref{ACS_FORSbandphot} shows this comparison.  A small shift is seen for all filters due to the difference in filter transmission. In conclusion, there are no significant deviations from a linear offset and we find that photometric errors are consistently smaller in the ACS photometry compared with the $BVI$ FORS1 photometry. Therefore, we chose our ACS photometry results throughout the subsequent analysis. 

\subsection{Completeness Tests}
\begin{figure}
\centering
\includegraphics[width=8.5cm]{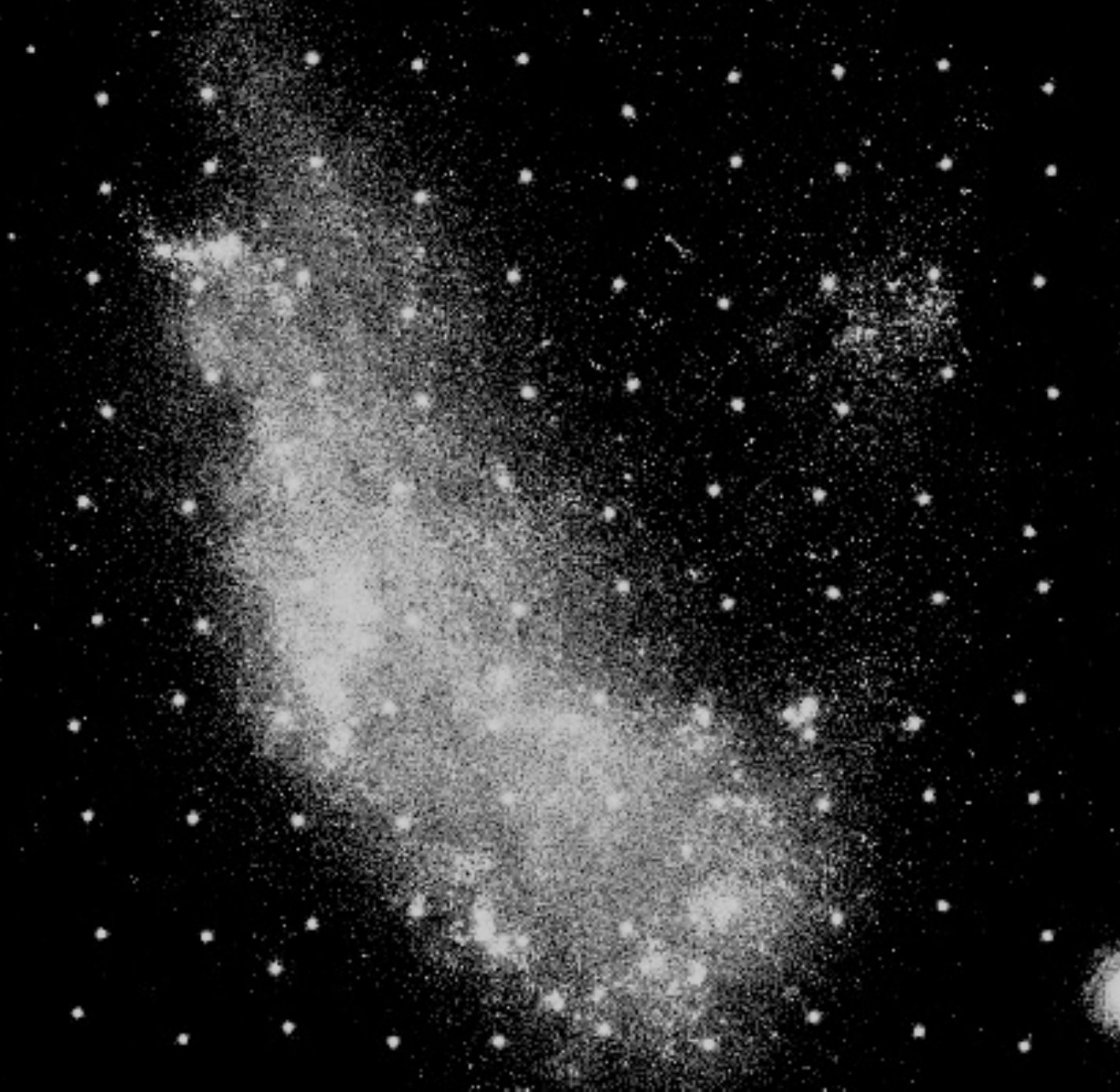}
\caption{Illustration of one test image realization where artificial star clusters are added to the F475W HST/ACS image of NGC\,1427A. We point out that the $10\!\times\!10$ artificial object grid includes random 0-100 pixel offsets for each realization.}
\label{completitud_imagen}
\end{figure}

Because we need to establish the photometric limits of reliability of measuring cluster magnitudes and sizes, we performed detailed completeness tests adding artificial clusters to the ACS images.~However, due to the importance of the U band for the mass and age determination, we do not apply any S/N cut to objects detected in the FORS1 U band.
A key point during this stage is to define the region where artificial objects are added to the images. Since we are aiming at studying the cluster populations across the whole galaxy, we perform our tests in the area of the main galaxy body, thereby considering the worst case scenario where objects are affected by the stellar crowding in the galaxy (see Figure~\ref{completitud_imagen}).

Our recipe consists of the following steps. First, we generate point-like and extended sources using ePSF library and  the {\sc Baolab} task {\sc MKCMPPSF} \citep{Ishape}. This task creates star cluster profiles by convolving the derived ePSF with a \cite{King} model with a concentration parameter $r_{\mathrm{tidal}}/r_{\mathrm{core}}=30$ with the desired FWHM. Second, we create a zero background image with artificial extended sources on it using the convolved ePSF as input for the {\sc MKSYNTH} task in {\sc Baolab}. For each magnitude and FWHM value one hundred artificial clusters are added in a $10\!\times\!10$ mosaic pattern with an initial distance of 200 pixels. A random shift from 0 up to 100 pixels is added in the $x$ and $y$ axis to each artificial object. Third, the zero background images are added to the science images. Finally, we run the photometry in the same way as it was done for the science frames. 

\begin{figure}
\centering
\includegraphics[width=8.9cm]{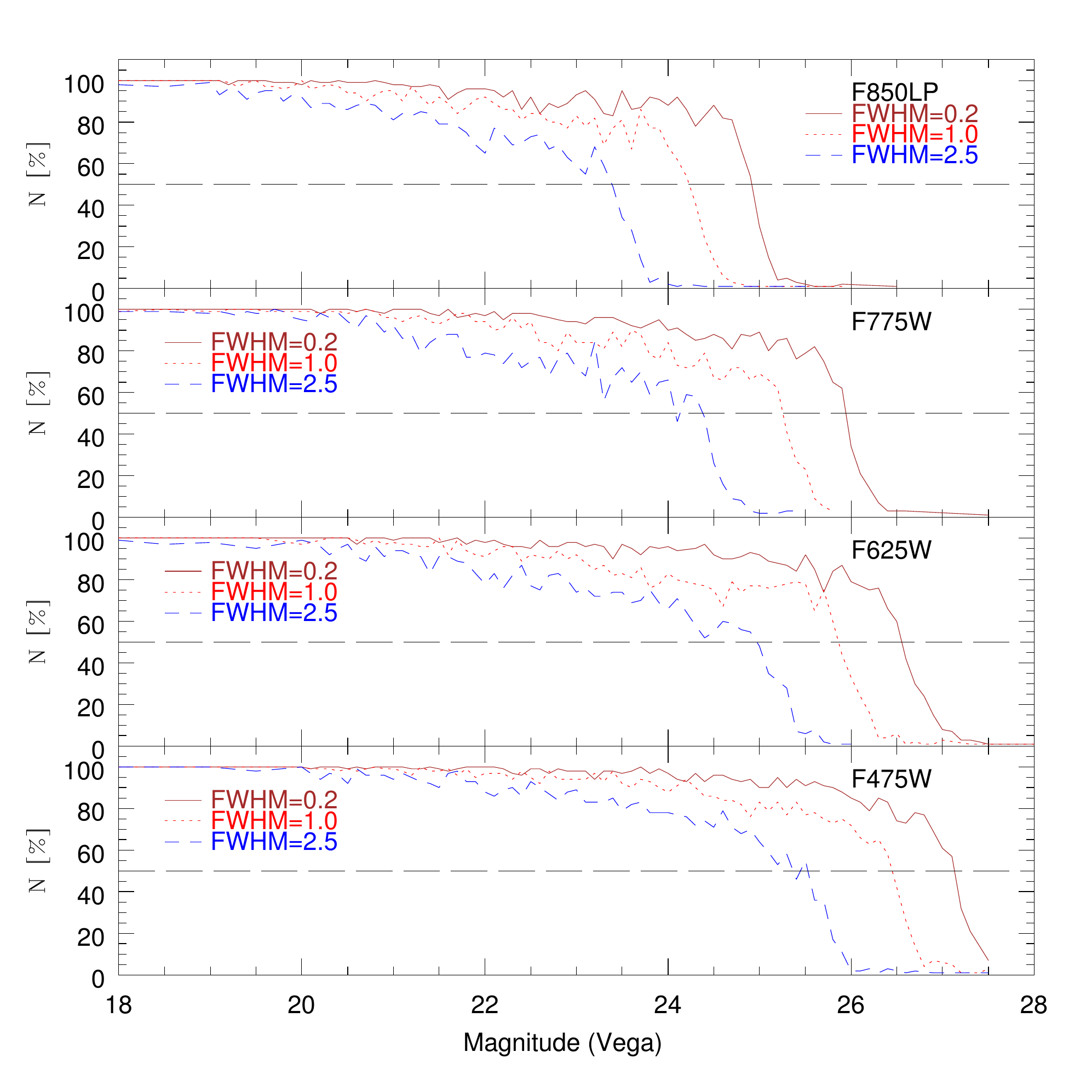}
\caption{Recovered number of objects as a function of magnitude and size.~In each panel, the corresponding ACS filters are indicated as well as the corresponding sizes of the simulated objects.~Horizontal dashed lines indicate the 50\% completeness limit.}
\label{completitud}
\end{figure}

\begin{figure}
\centering
\includegraphics[width=8.9cm]{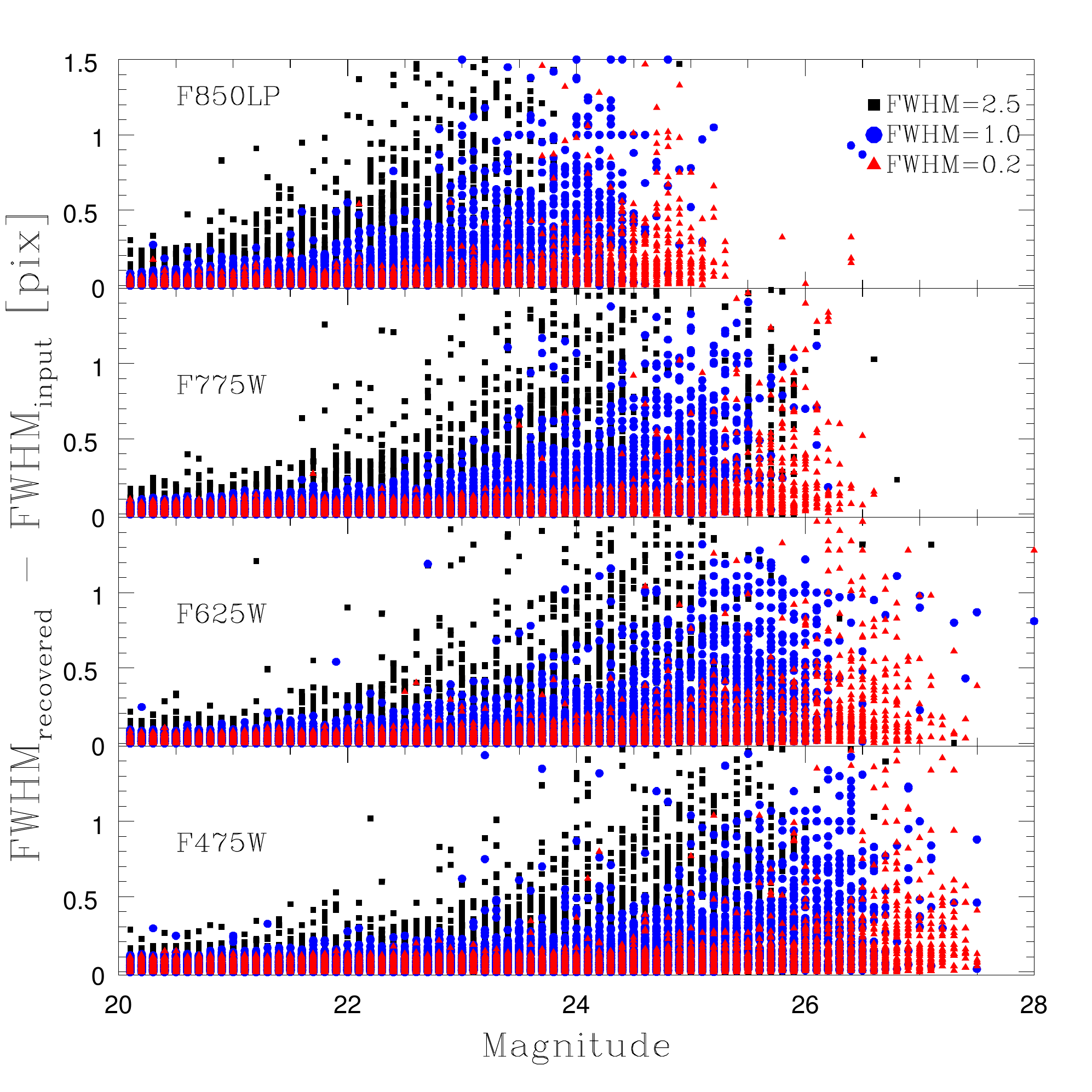}
\caption{Recovered sizes as function of magnitude for all ACS filters for objects with sizes FWHM=0.2 pix, 1.0 pix, and 2.5 pix.~The subsample numbers correspond to: I F475W, II F625W, III F775W, and IV F850LP and the sizes are expressed as the difference between the recovered and the input size. Vertical stripes correspond to the discrete magnitudes of the completeness tests. Horizontal stripes are a consequence of the numerical precision output of {\sc Ishape}.} 
\label{completitud_sizes}
\end{figure}

Completeness tests were performed for three FWHMs: 0.2, 1.0 and 2.5 pixels with the goal of studying a point-like object and two extended objects which, at the distance of NGC\,1427A, corresponds to $\sim\!0.8, 4$, and 10 pc, respectively. The results of our tests are shown in Figure~\ref{completitud} and \ref{completitud_sizes}.  

The number of recovered objects as a function of magnitude decreases faster for the extended ones than for point-like sources, which are affected by crowding at fainter magnitudes, as expected. The 50\% completeness limit is reached at F475W~$\simeq25.6$, F625W~$\simeq25.0$, F775W~$\simeq24.5$, and F850LP~$\simeq23.5$ mag for the extended objects (FWHM=2.5 pixels). 

Completeness tests also provide us with an estimate of the cluster size measurement error. Typically, bright and compact sources show an error in FWHM~$\!=\!0.1$ pixels  for all passbands beyond the 50\% completeness limit, while more extended ones (e.g. FWHM=1.0)  show the recovered FWHMs , increases toward fainter magnitudes. Because of the different filters resolutions and exposures,  size differences between the input and the recovered sizes are expected, and also increased, due to crowding effects (see Fig. \ref{completitud_sizes}).~Finally we adopted the following limiting magnitudes, consistent with our 50\% completeness and error selection criteria:  F475W~$\le\!25.6$, F625W~$\le\!25.0$, F775W~$\le\!24.3$, and F850LP~$\le\!23.0$ mag.

\subsection{Extinction Correction}
Finally, we take into account the absorption caused by the Milky Way interstellar medium that affects the magnitudes of each object. We adopted the extinction maps of \cite{2011ApJ...737..103S} which is a recalibration of the work by \cite{Schlegel1998}. Values where taken from NED\footnote{http://ned.ipac.caltech.edu\/} for each ACS filter considering  R$_{\mathrm{v}}=3.1$ and for the U band we adopted  the U CTIO bandpass.

\section{Results}
\subsection {Color-Magnitude Diagrams}

\begin{figure*}
\centering
\includegraphics[width=5.9cm]{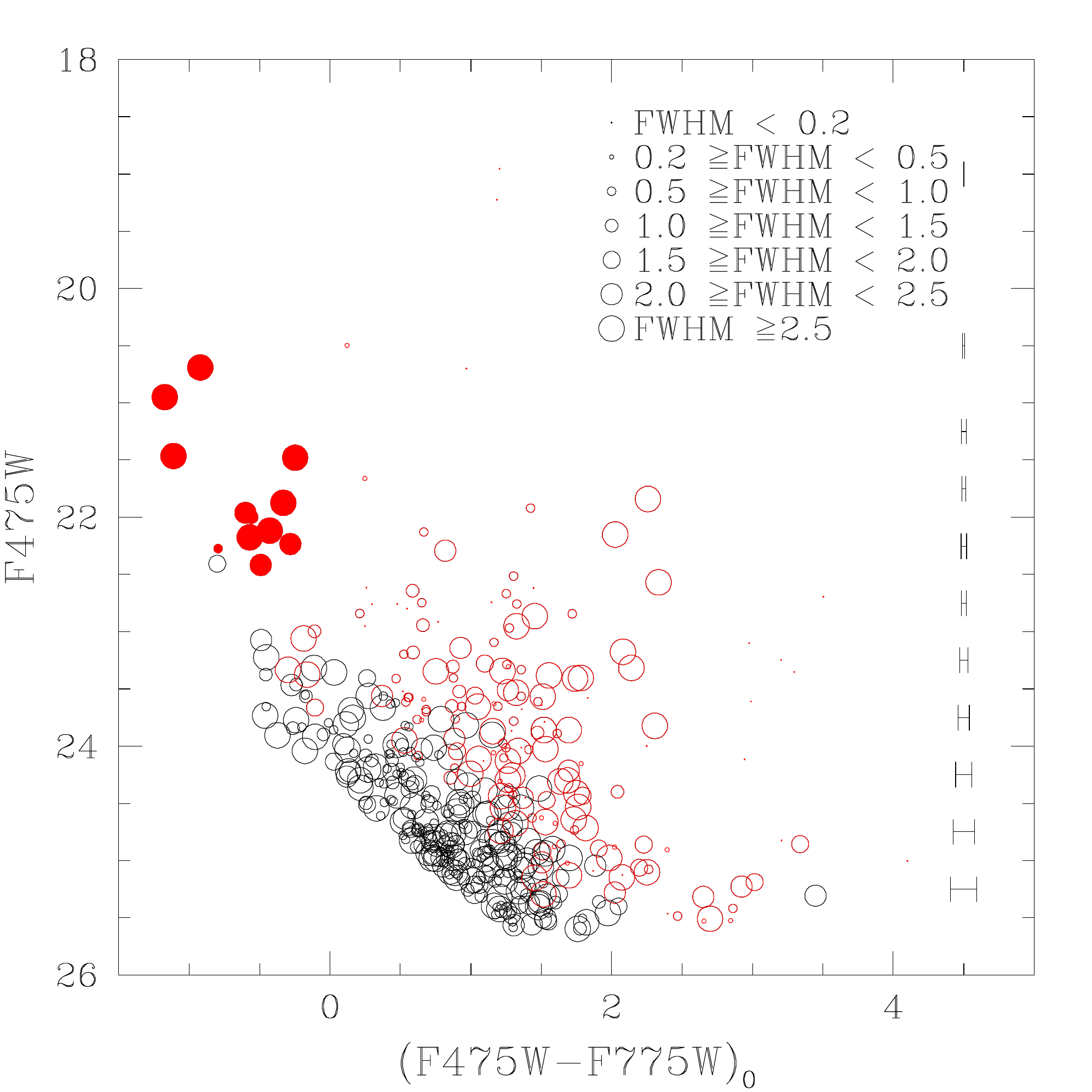}
\includegraphics[width=5.9cm]{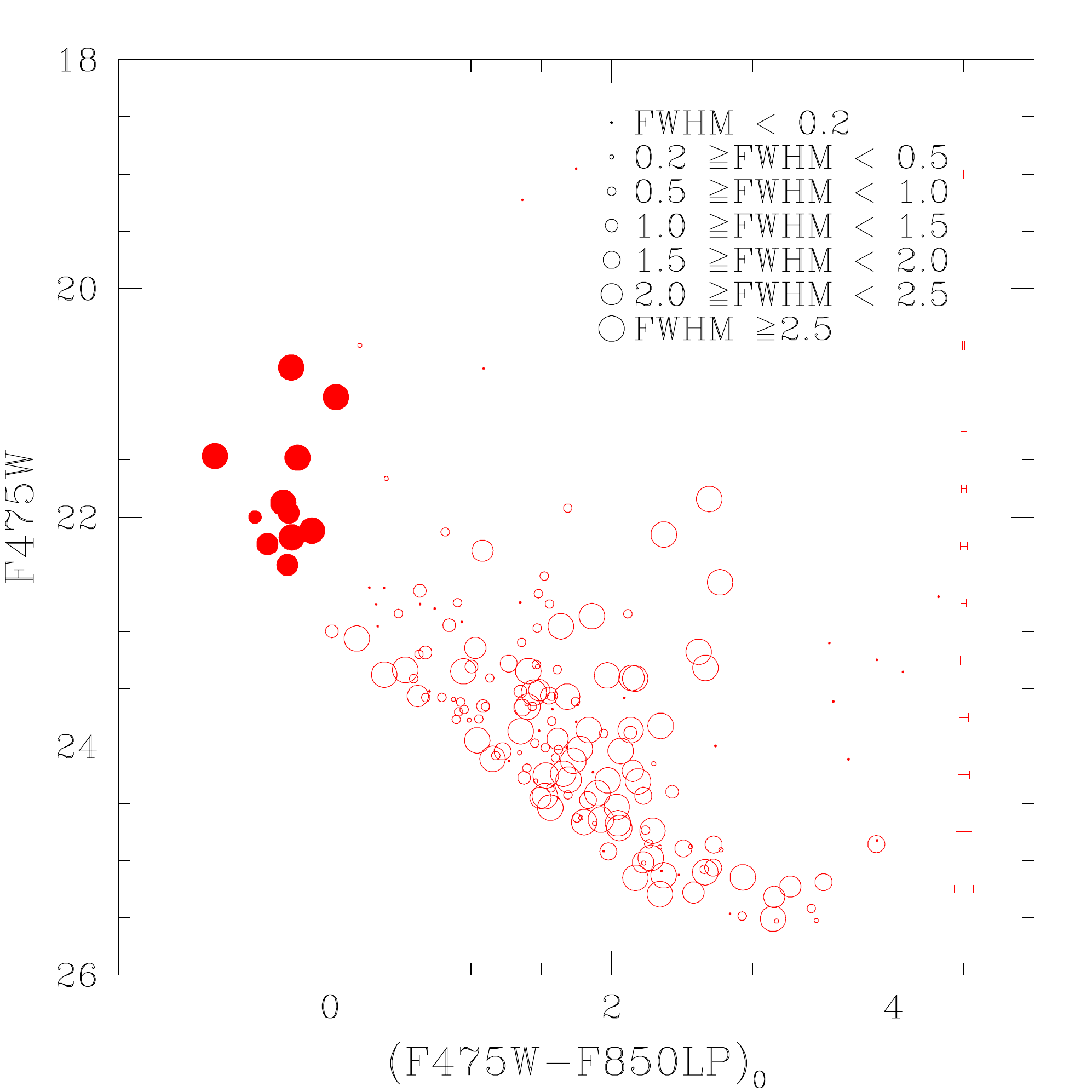}
\includegraphics[width=5.9cm]{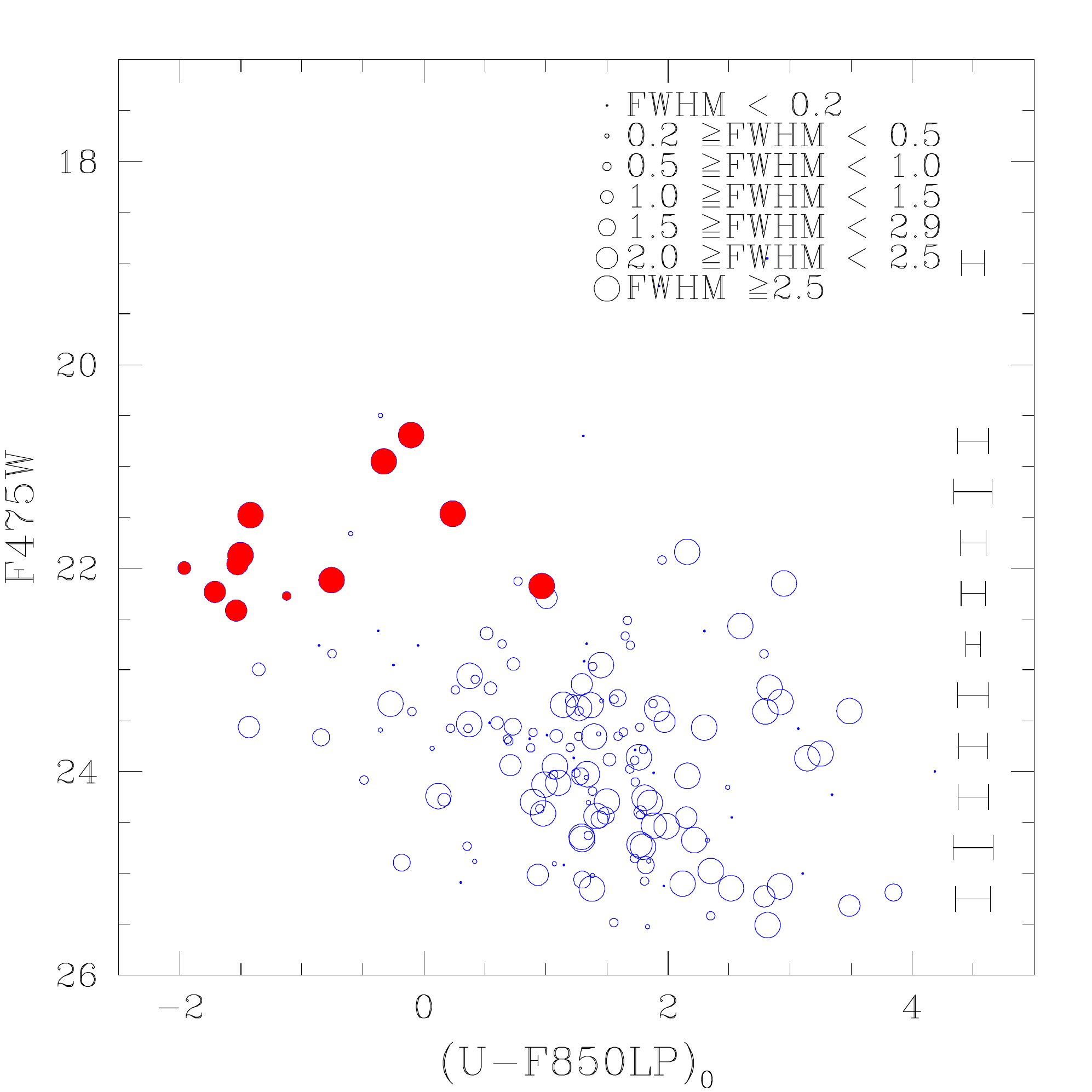}
\caption{Galactic extinction corrected color magnitude diagrams for sources in NGC\,1427A. In each panel, the FWHM values (in pixels) of the objects are parameterized by the symbol diameter that is proportional to the size measured in the F475W frame. Error bars located in the right-hand side of each panel represent the average photometric errors of the color axis at a given luminosity. In the case of the right panel, the average error bar is affected by the relatively large U-band errors and stochastic effects at bright magnitudes and does not shows the typical trend as function of magnitude. ({\it Left and center panel}): Black dots represent objects detected in three passbands (i.e.~F475W, F625W and F775W). Red open circles show objects detected in all four ACS passbands. Filled red circles correspond to the  star cluster complexes. ({\it Right panel}): Only objects that were detected in the all four ACS passband and the FORS U-band image are plotted in blue open circles. In this panel the young star cluster complexes are represented by red filled circles.} 
\label{VI}
\end{figure*}

In Figure~\ref{VI} we show the Galactic foreground  extinction corrected color-magnitude diagram (CMD) for all sources in NGC\,1427A. Plotted objects have ACS photometry errors $\leq\!0.1$ mag in each filter.%
~We do not impose a restriction in photometric errors for the U-band photometry due to its importance in mass-age determinations. In the left and center panel we find a distinctive group of luminous objects at F475W~$\le\!22.5$ mag and blue colors [(F475W-F775W)~$\lesssim\!0$]. The members of this group of star cluster candidates have clearly extended light profiles with FWHM~$\ge2.0$ pixels and larger in at least one ACS filter.~Visual inspection of the images shows that these objects correspond to extended star cluster complexes mostly located in the edges of NGC\,1427A. Therefore, using the left and central panels, we adopt the following selection criteria of  what will be considered as blue star cluster complex: objects with (F475W-F775W)$_0$ and (F475W-F850LP)$_0$ colors $\le\!0.2$ and F475W~$\le\!22.5$ mag.~With these criteria objects with errors larger than 0.1 mag will not appear in the CMDs. Because the observation acquired through the F475W filter are the deepest ones,
the number of detected objects decreases in the reddest passbands, and therefore we have objects, like the star cluster complex seen as black open circles in the left panel of Figure~\ref{VI}, which are not detected in the F850LP ACS filter.

The right panel of Figure~\ref{VI} shows the F475W versus U-F850LP CMD composed of the deepest photometric band and a color with the widest spectral energy distribution (SED) coverage.~In this panel, the star cluster complexes appear less clustered with a wider spread in color than in the previous panels. Part of this spread may be due to the poorer photometric quality of the FORS1 U-band images and, therefore, the star cluster complexes may appear blended or contaminated by nearby objects. We have visually checked our images for whether this is the case for all these sources and find that {six sources are blended.~However, due to their location in the galaxy, we cannot discard nearby object contamination, nor galaxy environment contamination.}~Another possibility {{(since object magnitudes are corrected by Galactic extinction)}} for the larger color spread is the internal NGC 1427A extinction, but our SED fitting analysis shows that the derived individual star cluster complex extinction values (see Section~\ref{ln:AME}) cannot explain the spread seen in this panel. We are therefore left with the hypothesis that the color spread indeed implies a spread in the stellar population properties of these star cluster complexes.

\subsection{Color-Color Diagram}
\begin{figure}
\centering
\includegraphics[width=8.9cm]{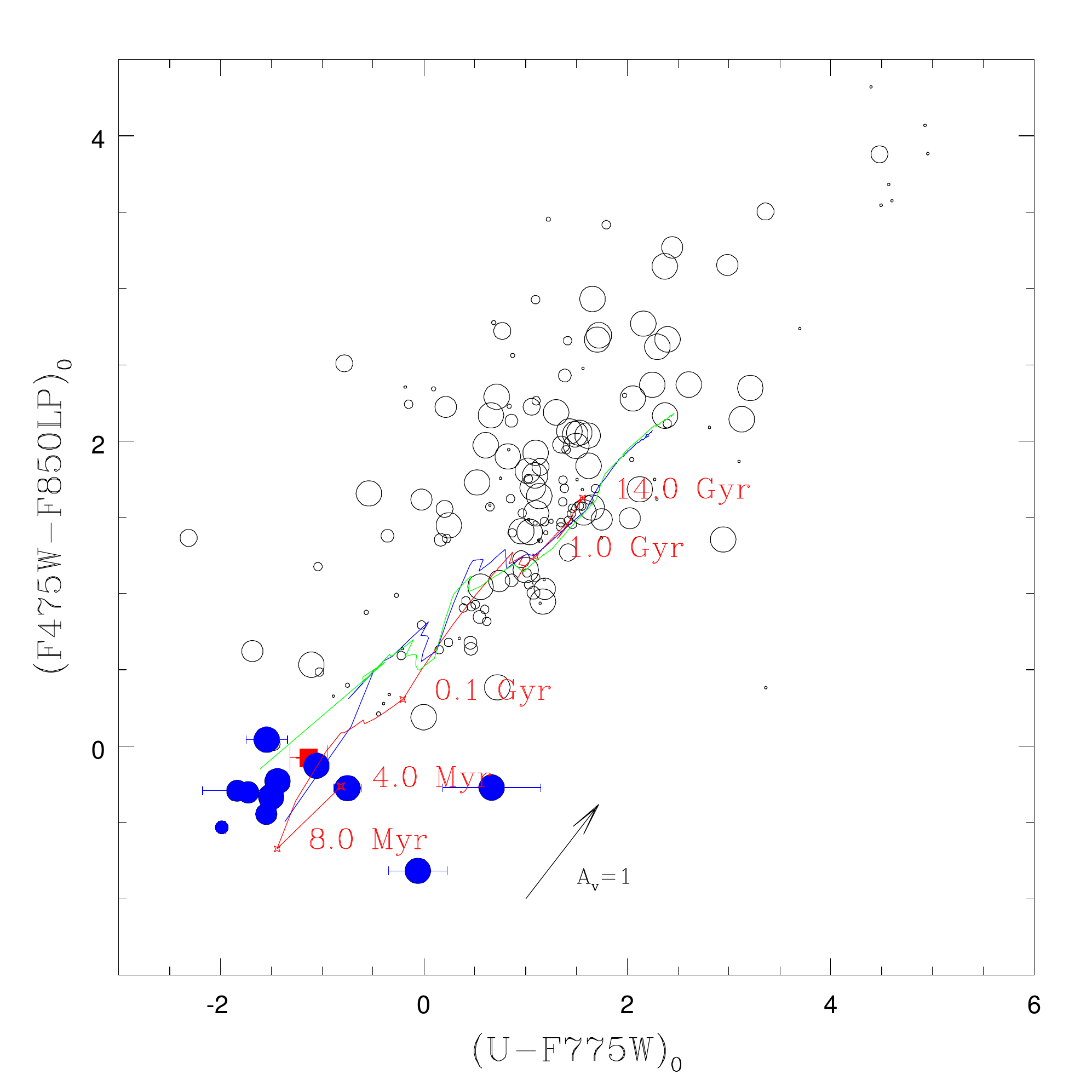}
\caption{Color-color diagram for sources in NGC\,1427A. Blue filled dots correspond to the star cluster complexes and the red filled square corresponds to the luminosity-weighted averaged color of the star cluster complexes. Overplotted lines are SSP models from GALEV \citep{GALEV} for metallicities $Z\!=\!0.02$ (red), $Z\!=\!0.008$ (blue) and $Z\!=\!0.004$ (green). Red labels indicate the stellar population age for the solar metallicity track. The arrow shows the color displacement for an object with an intrinsic extinction of $A_V\!=\!1$ mag. Photometric errors are only visible in (U-F775W)$_0$ color because the (F475W-F850W)$_0$ error bar is smaller than the symbol size.}
\label{gz}
\end{figure}

Figure~\ref{gz} shows the (U-F775W)$_0$ versus ~(F475W-F850LP)$_0$ color-color diagram that combines the most homogeneous SED coverage {(i.e.,~objects detected in five passbands)} of our photometric data into a diagnostic diagram from which we can infer rough stellar population ages and metallicities of the blue star cluster complexes.~From this diagram, it is clear that the spread in colors as seen in the right panel of Figure~\ref{VI} cannot be solely attributed to photometric errors and internal  {{NGC 1427A extinction (or the line of sight) corrected by the Galactic extinction law}}, 
considering the direction of the extinction vector and the individual photometric errors. 
To make our analysis more comprehensive we show the position of a fiducial star cluster complex (filled red square) considering the individual magnitudes of the star cluster complexes, assuming that star cluster complexes are coeval.

We overplot SSP model predictions from the GALEV \citep{GALEV} for stellar population metallicities $Z\!=\!0.02$, 0.008, and 0.004, and ages 3 Myr to 14 Gyr. In this diagram, our star cluster complexes lie at the bottom left side of the panel at (F475W-F850LP)$_0\!\simeq\!-0.3$ mag indicating that they have relatively young ages of a few Myr. The red filled square (i.e. the fiducial star cluster complex) falls near the SSP models of solar metallicity and $Z=0.008$ at the stellar population age of $\sim\!4$ million years. Considering all objects in our cluster candidate sample and judging from the position of the bluest star cluster complexes in the color-color diagram, we conclude that we are observing the youngest star cluster complexes in NGC\,1427A. Their magnitudes are listed in the last six columns of Table A1. 

\subsection{Determination of Cluster Age, Mass and Intrinsic Extinction}
\label{ln:AME}
\begin{figure*}
\centering
\includegraphics[width=8.9cm]{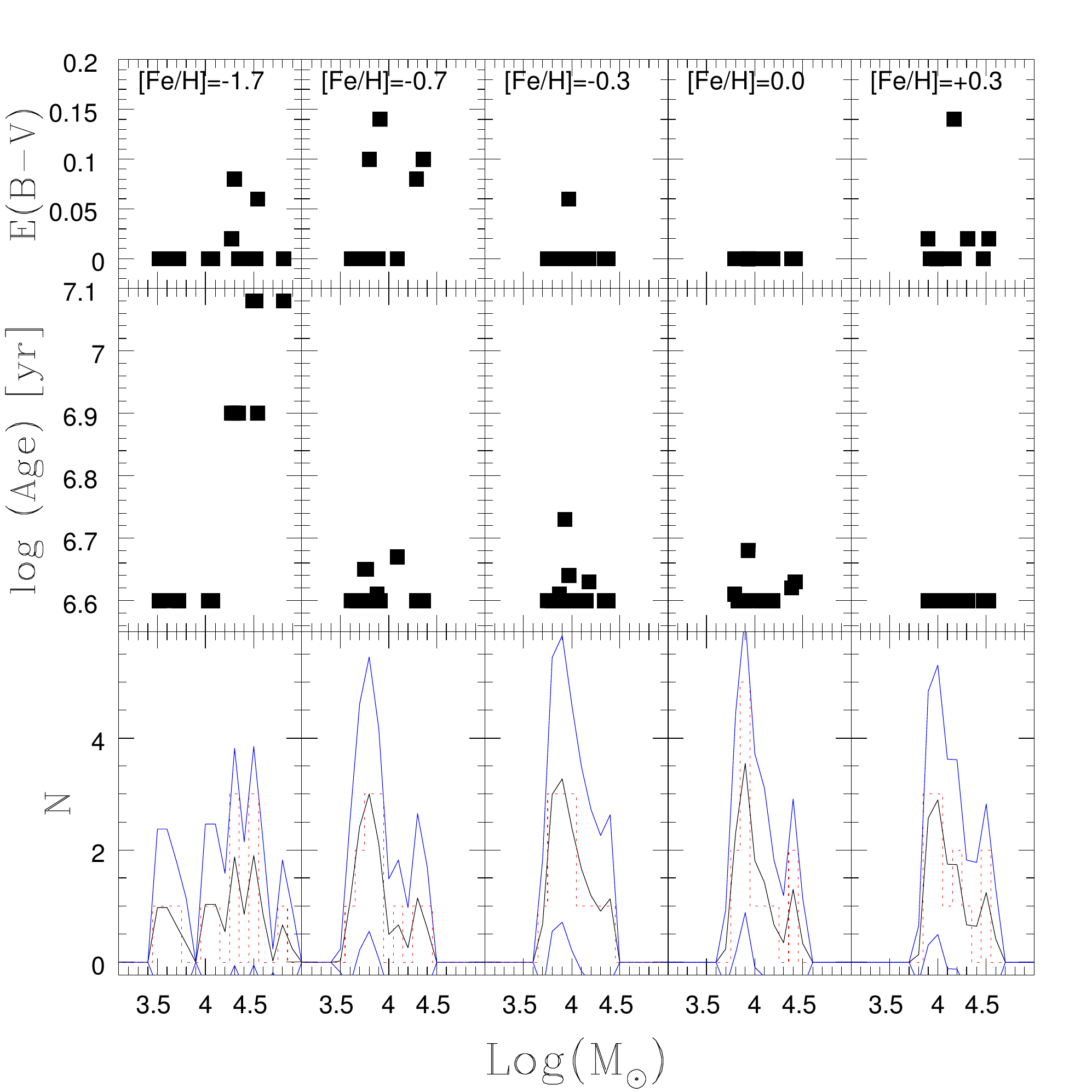}
\includegraphics[width=8.9cm]{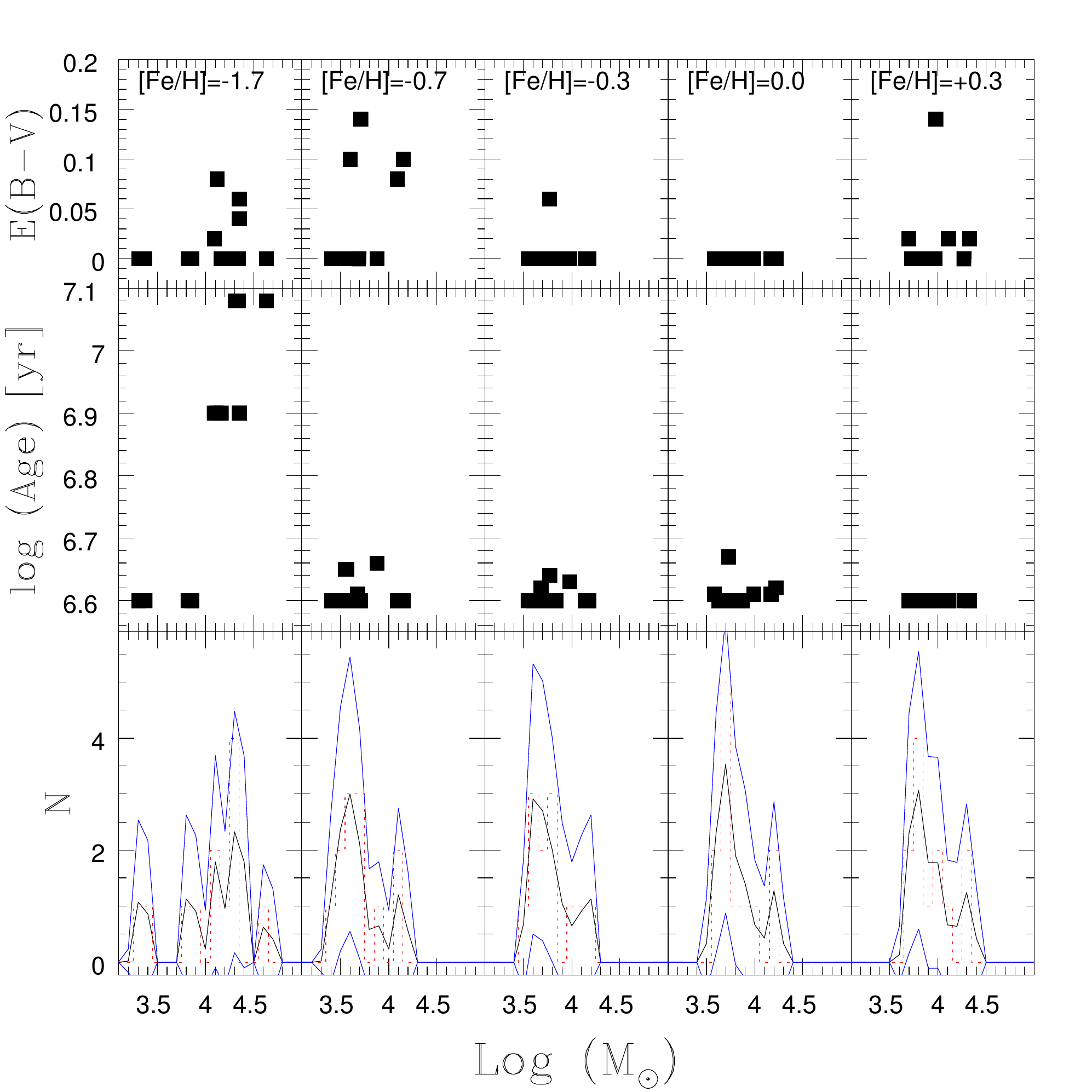}
\caption{Correlation plots for stellar population ages, masses and total extinction values of the blue star cluster complexes, derived from SED fitting using GALEV SSP model predictions \citep{GALEV} for a Salpeter IMF {({\it left panels})} and Kroupa IMF {({\it right panels})}.~For both plots, the top-row panels show extinction-mass correlation plots for stellar population metallicities [Fe/H]~$=-1.7, -0.7 -0.3, 0.0,$ and $+0.3$ dex. The center-row panels illustrate the correlation between stellar population age and mass, while the bottom-row panels show the resulting stellar mass distributions of the star cluster complexes for the corresponding five stellar population metallicities. {The histogram (in red dashed lines) bin sizes in the bottom panels are 0.1 dex in $\log({\cal M}_\odot)$ while the solid lines correspond to a Gaussian kernel density estimator of width half the bin's size,} and their 90\% confidence limits shown in blue. Individual values are listed in Table A2.}
\label{age_mass}
\end{figure*}

In the following we are using the photometry from all five passbands to estimate the stellar population ages, masses and extinction values of each blue star cluster complex via SED fitting. For this we compare the star cluster complex SEDs to GALEV SSP model predictions \citep{GALEV}, considering {{the two cases}} of initial mass function (IMF): a  \citet{Salpeter:1955} IMF with a stellar mass range from 0.1 up to 100 ${\cal M}_\odot$ and a \citet{Kroupa:2001fk} IMF with a range in stellar mass from 0.1 up to 100 ${\cal M}_{\odot}$ with a broken power-law of index 1.3 for $0.1\le {\cal M}/{\cal M}_{\odot} \le0.5$ and a power index of 2.3 for ${\cal M}/{\cal M}_{\odot}\!\ge\!2.5$.~The GALEV model calculations include full emission lines
treatment for young stellar populations,\footnote{GALEV models consider the continuum, which in combination with the emission lines, can account  up to $50\%-60\%$ of the emitted flux for young stellar population at low metallicities in broad band passbands. {{Ionizing  photons are taken from  the tabulated values from \citet{Schaerer:1997vn}. For the N$_{LyC}$ at youngest ages the fraction correspond to:  N$_{LyC}$=51.343~(Solar), 51.765~(Z$=-0.3$), 51.919~(Z$=-0.7$), 51.850~(Z$=-1.7$), 51.109~(Z$=+0.3$) for Kroupa and N$_{LyC}$=51.143~(Solar), 51.561~(Z$=-0.3$), 51.715 (Z$=-0.7$), 51.860~(Z$=-1.7$), 50.907~(Z$=+0.3$) for Salpeter}}  \citep{Anders:2003qy}. } i.e.,~the models include bound-free emission from hydrogen and helium, free-free emission and emission from the hydrogen two-photon process as well as other less prominent line emissions. 
Stellar population age, mass and extinction toward the star cluster complexes is derived through a least-square fitting technique following \citet{Bik:2003} and \citet{Anders:2004SED}. 
In Figure~\ref{age_mass} we show the derived stellar population ages, masses, and total extinction values for the star cluster complexes.  {{Each measurement is plotted against mass.}} Derived extinctions are typically $E(B\!-\!V)\!\le\!0.15$ mag for all models. However, most objects do not show any significant extinction values for all model parameters, {{nor differences between extinctions at low and high masses, }} which leads us to estimate the average extinction towards the complexes as $E(B\!-\!V)\!=\!0.03$ mag. The ages of most star cluster complexes are consistent with objets younger than 3.9 million years, while four star cluster complexes show ages of $\sim$ 3 million years older than the others for models with solar to slightly sub-solar metallicities. Therefore, the main conclusion is that we are detecting the youngest star cluster formation sites {{(i.e. star cluster complexes)}} in the galaxy. Derived masses of the star cluster complexes show a range between $\sim\!2500\,{\cal M}_{\odot}$ up to $3.1\!\times\!10^{4}\,{\cal M}_{\odot}$ depending somewhat on metallicity.~This classic mass-metallicity degeneracy is seen as a trend of increasing stellar mass for higher metallicities.
{{In order to compare how the derived masses distribute for each metallicity and IMF, a {{Gaussian kernel density estimator with bandwidth of the bin size}} was applied to the mass histogram of the  Figure ~\ref{age_mass}. Identical metallicities and different IMFs, show histograms where mass distribute  in a similar way and the shape of the mass histograms tend to be similar for all four metallicities (except for Fe/H=-1.7), being somewhat consistent among themselves as a group. }}
Nevertheless, the method used in the determination of the mass distributions of these star cluster complexes should not cover the fact that all derived masses are in the regime of significant influence of stochastic variations in the constituent stellar populations \citep[e.g.][]{Cervino:2004fk, Fouesneau:2010uq, Fouesneau12}. %
~Future adaptive-optics supported observations at near-IR wavelengths will shed more light on this particular aspect and the accuracy of our star cluster complex mass determination.

{Furthermore, due to the limitations of SSP models and the accuracy of our photometry, we cannot conclude whether the star cluster complex formation started at the same time across NGC\,1427A or whether there is a spatial correlation of star cluster complex ages. It will be interesting to further investigate the remaining young star cluster population using spectroscopic observations (to be presented in a forthcoming paper) and to derive more accurate ages and address the question of spatially correlated star formation.}

\subsection{Determination of Star Cluster Sizes}
\begin{figure}
\centering
\includegraphics[width=8.9cm]{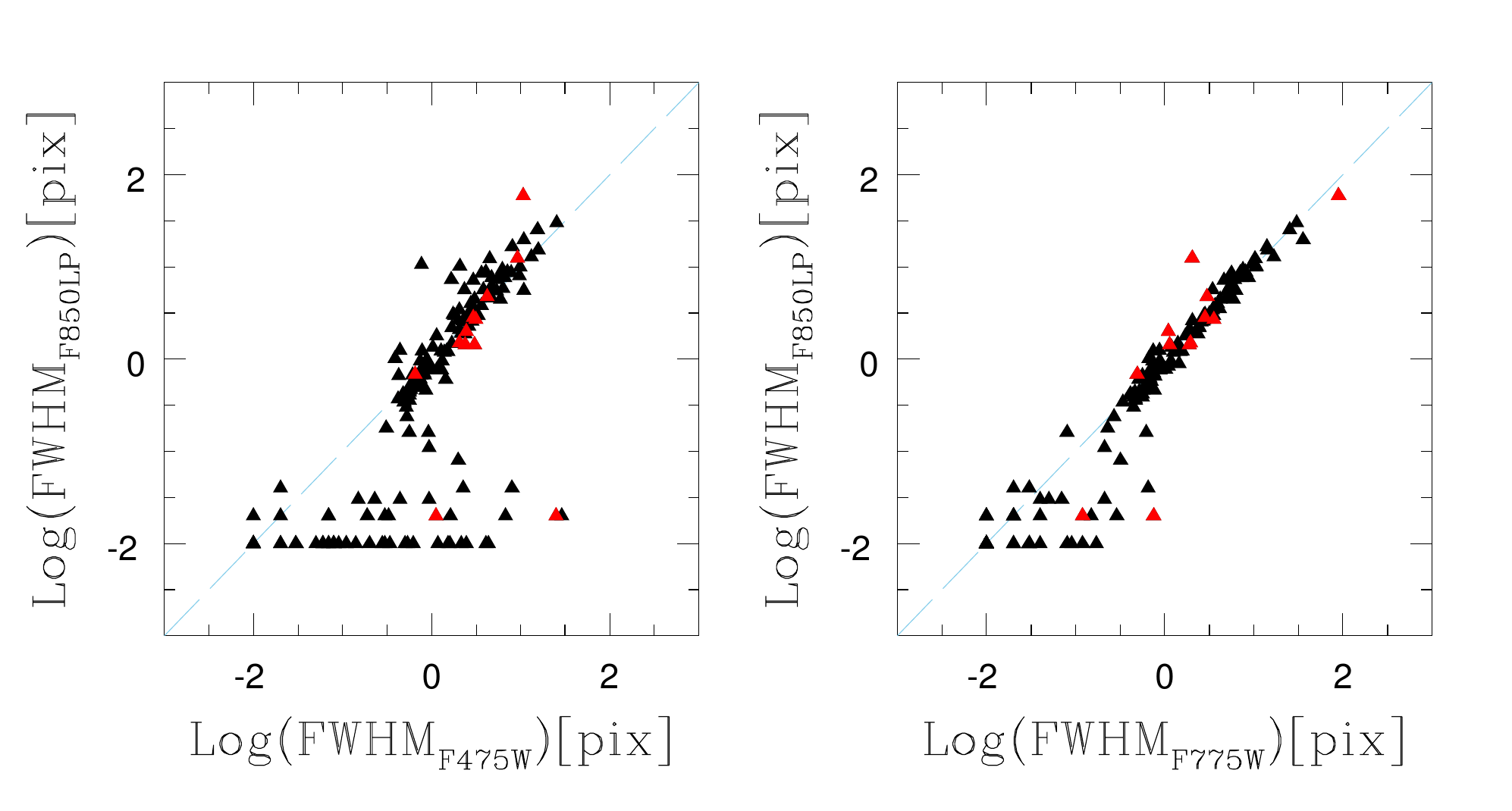}
\caption{Size comparison plots for objects detected in NGC\,1427A. Solid triangles indicate individual star cluster candidates that fulfilled our selection criteria, while the red-labeled data correspond to the blue star cluster complexes. The dashed line corresponds to the one-to-one relation.} 
\label{filter_sizes}
\end{figure}

The sizes of all sample objects were measured in all ACS passbands using {\sc Ishape} \citep{Ishape} as described in Section~\ref{ln:photsize}. In general, object sizes should be consistent among the different passbands and reflect the wavelength dependent resolution limits of each passband. In Figure~\ref{filter_sizes} we compare the measured FWHM values in two filter against the corresponding measurements in selected filters. The sizes of the star cluster complexes are indicates as red labels 
listed in Table~\ref{tablefinal}. To guide the eye, we add to each panel a dashed line that represents the one-to-one relation. It is clearly seen that most objects lie on or close to the one-to-one relation, fully consistent with the measurement uncertainties. {The measured dispersion above 1 pixel resolution in the left panel corresponds to 0.54 and in the right panel to 0.02, difference due to the proximity in wavelength of the selected ACS filters.}  The resolution limit of our data can be seen as the relatively sudden increase in scatter below a few sub-pixels. However, few resolved objects above this limit do not follow the one-to one line and the size differences cannot be explained as a result of the changing resolution limits of various filters.~For example, two star cluster complexes (ID:6002 and ID:3070) are extended in the F475W images, but show point-source like sizes in the F850LP filter.

As it is expected, star cluster complexes are brighter in bluer passbands than in red passbands due to their young ages. However, at such young ages (i.e.~few Myr) the star cluster complexes appearance and therefore size is dependent on the existence of blue and red supergiant stars. From our size comparison exercise, we conclude that few star cluster complexes in our sample show indications of simultaneously hosting blue and red supergiant stars in different locations within their star cluster complex extent, thereby producing the observed size differences. However, given the expected stochasticity of the constituent stellar populations for the observed total stellar mass, it is surprising that not more star cluster complexes show such size offsets. This, in turn, suggests that most star cluster complexes are already very well mixed in their stellar content at such young ages.

\subsection{The Star  Formation Rate}
In the following we compare the star-formation rate (SFR), derived from the total H$\alpha$ flux, with the star-cluster formation rate (SCFR) to derive a star-cluster formation efficiency in NGC\,1427A.  The velocity of NGC\,1427A %
shifts the H$\alpha$ line emission into the F660N filter. 
{{We assume that the galaxy's H$_{\alpha}$ flux resides inside an aperture with a radius of $\sim$ 6 kpc.}}   
In order to measure the H$\alpha$-based SFR, we reduced the ACS F660N images following the same procedures as for the other broad-band ACS images. Prior to the photometry the ACS F660N image was multiplied by the photflam header value in order to obtain fluxes in units of erg~cm$^{-2}$ \AA$^{-1}$. We masked background objects and performed aperture photometry centered on the galaxy with a radius of 1500 pixels ($\sim\!5.97$ kpc) with an annulus of 300 pixel ($\sim\!1.19$ kpc) width in which the background/sky flux was measured.  {{Since it is uncertain where the galaxy surface brightness is equivalent to the sky surface brightness, we provide a range of values derived from our data to quantify this measurement uncertainty. For this purpose,}} among all sky/background measured we adopted the highest and lowest values  $\mu=23.684$ and $\mu=22.806$ mag arcsec$^{-2}$.

Finally, we derived an averaged H$\alpha$ flux of: 
\begin{equation}
F_{\mathrm{H}\alpha}=2.2623 \pm 1.237 \times 10^{-14}\,\,  \mathrm{erg ~cm}^{-2}\mathrm{\AA}^{-1}
\end{equation} 
Where the errors correspond to the H$\alpha$ flux considering the maximum and minimum measured background/sky values. 
In view of the F660N passband includes the [N{\sc ii}] 6548, 6583 \AA\ emission line-doublet, we estimate the [N{\sc ii}] contamination by adopting the solar H/N ratio, which yields a line flux ratio of $\sim\!1.689\!\times\!10^{-4}$. The [N{\sc ii}]- corrected $F_{\mathrm{H}\alpha}$ value is then:  
\begin{equation}
F_{\mathrm{H}\alpha}=(2.2619\pm1.236) \times 10^{-14}\,\, \mathrm{erg ~cm}^{-2}\mathrm{\AA}^{-1}
\end{equation}
The current SFR can then be derived using the relation from \cite{Hunter:2004fk}:
\begin{equation}
\dot{{\cal M}}=5.96 \times 10^{-42} L_{{\rm H}\alpha} 10^{0.4A_{{\rm H}\alpha}}\,\, [{\cal M}_\odot\, {\rm yr}^{-1}]
\end{equation}

where L$_{{\rm H}\alpha}$ is the H$\alpha$ flux in erg s$^{-1}$.~We adopt the extinction coefficient value from \cite{Georgiev2006}, i.e.~$A_{{\rm H}\alpha}=0.811A_V$, and a distance to NGC\,1427A of 19 Mpc. Finally, we derive the current SFR as $\dot{{\cal M}}\!=\!0.05\!\pm\!0.03\, {\cal M}_{\odot}\, {\rm yr}^{-1}$. This value agrees with the one previously reported by \cite{Georgiev2006}.

\subsection{The Star Cluster Formation Rate}
{{Since the masses of star cluster complexes are derived using models for a range of metallicities and different IMFs, we will calculate the SCFR accounting for this spectrum of possibilities.

Some caveats need to be accounted for.  First, ages  of the models are provided   at intervals of 4 million years starting from 4 million years,
therefore ages reported to be within those intervals have been obtained by interpolation.  Moreover, if a star cluster complex is assigned with the youngest age from the model, it is likely that this object is younger than the age available from the model.

We compute the range of SCFR allowed by the  uncertainties in our mass, age, and extinction fits derived from the SSP models as described in \S\,3.3 and listed in Table \ref{tablefinal2}.  The resulting SCFRs are listed in Table \ref{TABLA_GAMMA}  for representative combinations of IMF and metallicity.
}}
 {{ Certainly, another source of uncertainty that affects any mass-age determinations is related to the ingredients of the models used for the analysis.  For example, as  found by \cite{Calzetti:2010ve} and \cite{Andrews:2013ly}, a change  in the  upper mass limit of the IMF  
 will  change the mass estimation by a factor of 2.5. Nevertheless masses are  expected to be within the order of magnitude
 (as will be seen ahead, in Table  \ref{TABLA_MASS_ERROR}). }}
 {However, we cannot ignore that despite the consistency of our results for the two IMFs, we are in a lower mass regime, which has an impact in the derived masses, due to the stochasticity that underestimates the masses up to 0.5 dex \citep[see Fig. 3 from][]{Fouesneau:2010uq}. }%
{{A second possible caveats could be an age spread of 10 Myr, which is roughly the oldest  age derived for the complexes, however all  star cluster complexes included in our relevant measurements show detectable flux in H${\alpha}$ and ages consistent with the time needed  to form a complete population of clusters, which is similar to 10 Myr \citep{Maschberger:2007uq}. 
Moreover, since we are dealing with an  unique episode of star formation,  we have assumed that the IMF does not change during this episode.  In \S\,3.7 we will compute the fraction $\Gamma$ of star formation occurring in clusters to that of the entire galaxy, and it will be calculated over this 10 Myr period, thus unlikely to be highly sensitive to the SSP assumption, but not free from stochasticity.  This, however, will not be the case for the IMF assumption (see \S\,3.7).}}

\begin{figure}
\centering
\includegraphics[width=8.9cm]{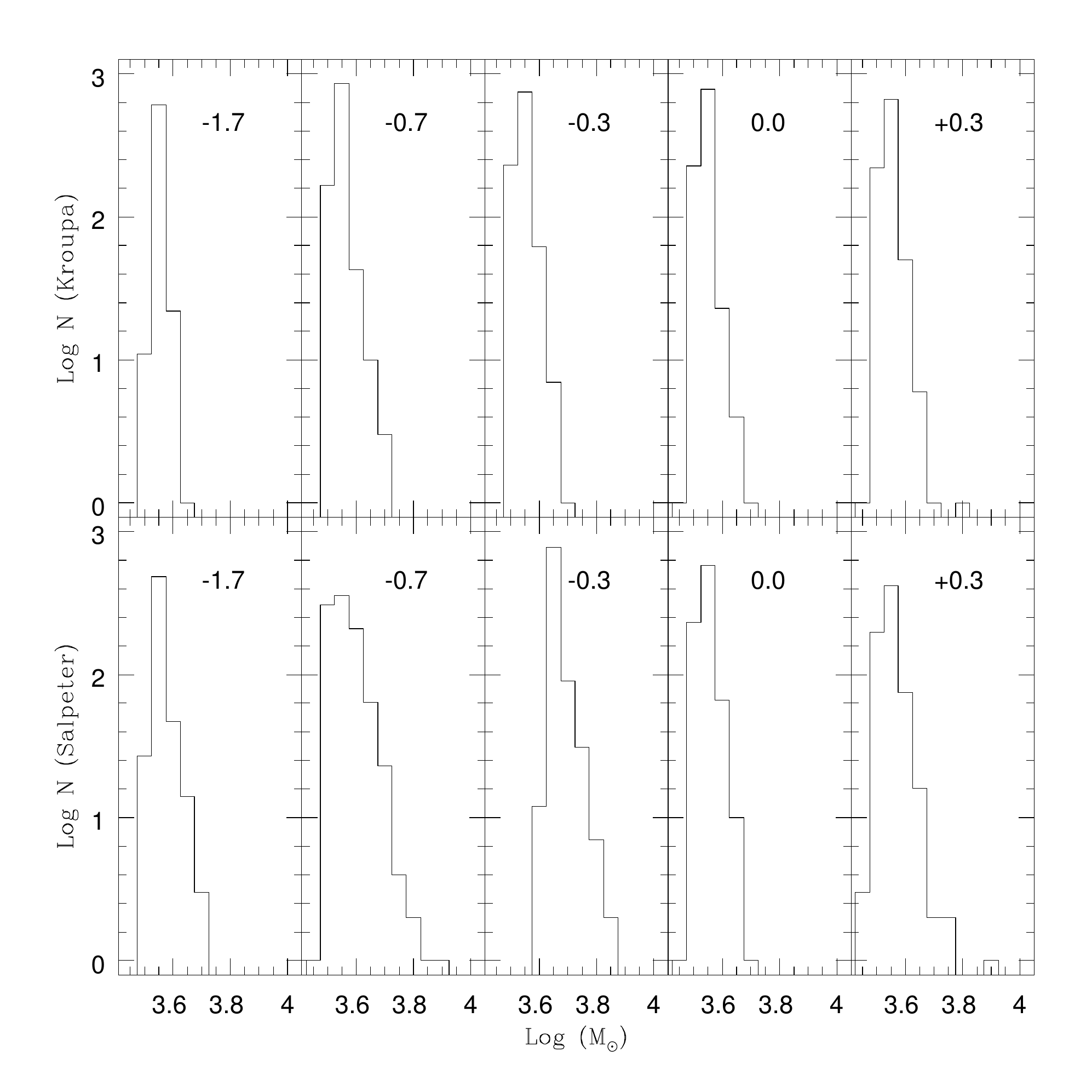}
\caption{Histograms of recovered masses for  objects with an input of FWHM=2.5 pix and solar mass corresponding to $10^{3.6}$ M$_\odot$. Bin sizes correspond to 0.05 in $\log (M_\odot)$ units.  
The upper panels correspond to Kroupa IMF and lower panels correspond to Salpeter IMF. Metallicity is listed on each sub panel.}%
\label{recovered_mass}
\end{figure}

{{Moreover, since we have a few cluster complexes with sizes larger than 50 pc,}} 
one might worry that this SCFR determination could
be biased to higher values, due to the contribution of individual
older and smaller star clusters blending into our star cluster
complexes (be it via projection effects or by being marginally
resolved/unresolved) or even field contamination {{ and thus biasing our measurements of
$\Gamma$ in \S\, 3.7 toward higher values. 
Ideally one can derive the typical disruption time \citep[e.g.][]{Boutloukos:2003qf,Lamers:2005ty,Whitmore:2007yq} and  the corresponding "infant mortality"  \citep{LadaLada},  and then estimate the fraction of the measured mass that may be due  to field stars.  However, since our sample is magnitude limited (dominated by the U-band) we cannot extrapolate any disruption scenario because we are analysing complexes with the youngest ages provided by the models. Moreover, we also do not know the fraction of embedded vs open star clusters in NGC1427A that can gives insights of the mass fraction coming from cluster and from field stars.

 In order to quantify the field and intra cluster contamination, we have simulated across 
  our image 1296 artificial objects in a grid of 36 $\times$ 36  objects separated by 100 pixels each and  centred in the galaxy and its surrounding following the same procedure as we did during the completeness tests. Our aim is that artificial objects will be affected by its local position on the image and thus affected by crowding and objects (stars, clusters, etc) located under the artificial object, giving us an empirical approach to the contamination that affects our complexes. Because the derived complexes ages correspond to the youngest age from the model, we have simulated, for each metallicity, artificial objects with magnitudes corresponding to the youngest age.  In order to be consistent with the completeness tests we have considered objects with a FWHM=2.5 pix.
We would like to mention that the derived masses correspond to luminous masses and GALEV models are normalised to $10^6$ M$_{\odot}$. Therefore,  we have magnitudes of the artificial objects are escalated to the magnitudes corresponding  to$10^{3.6}$M$_{\odot}$  which is the lowest mass derived for the complexes and therefore most likely to be affected by contamination. 

Detected magnitudes from artificial object were plugged into our fitting routine and recorded the output masses.  
In Fig \ref{recovered_mass} we show histograms of the recovered masses for each metallicity. Overall, input masses were recovered, but there is a spread due to the local variations (crowding, local contamination, etc). The most extreme recovered values are shown in table \ref{TABLA_MASS_ERROR}, allowing us to conclude that  the larger source of error is  due to the  
mass overestimation (up to 23\%)  rather than from mass underestimation (down to 4\%). 
}}

\begin{deluxetable}{ccccc}[!ht]
\tabletypesize{\scriptsize}
\tablecaption{Mass variation from artificial tests  \label{TABLA_MASS_ERROR}}
\tablewidth{0pt}
\tablehead{ 
\colhead{[Fe/H]} & \colhead{M$_{\mathrm{low}}$} & \colhead{M$_{\mathrm{high}}$} & Low \%& High \%
}
\startdata
\hline
 &&Kroupa &&\\
\hline
     -1.7   &    3.51  &  4.41  & 2.5&22.5\\ 
    -0.7   &     3.47  &  3.73  & 2.6& 3.6\\ 
    -0.3   &     3.48  &  3.66  & 3.3&1.7\\  
      0      &     3.47  &  3.70  & 3.6&2.8\\ 
     0.3    &     3.46  &  3.69  & 3.9&2.5\\ 
\hline
& &Salpeter& &\\
\hline
     -1.7   &  3.51 &   4.38   & 2.5&21.7\\ 
     -0.7   &  3.46  &  3.83   & 3.9&6.4\\ 
     -0.3   &  3.61  &  4.17   & 0.3&15.9\\ 
      0      &   3.46  & 4.33    & 3.9&20.3\\ 
      0.3    &  3.46  & 4.26    & 3.9&18.3    
\enddata
\tablecomments{M$_{\mathrm{low}}$ and M$_{\mathrm{high}}$ corresponds to the lowest  and highest recovered value in $\log{\mathrm{M}_\odot}$. Low \% and High \% corresponds to the  difference (in percentage) between the input mass and lowest/highest recovered value. }
\end{deluxetable}

 {{In conclusion, we do not know the fraction of intra cluster stars and field contamination, but this fraction should not bias our result to higher $\Gamma$ values more than the uncertainties estimated to be of the order of 22.5\%}, as will be seen below.
 Finally, in the last column of Table \ref{TABLA_GAMMA}, we present the final result, corrected for contamination ($\Gamma_{corr}$).}

\subsection{Fraction of Recent Star Formation occurring in Clusters}

The above range in SCFR then translates into a corresponding range for the fraction of star formation occurring in star clusters, introduced by Bastian (2008) as $\Gamma =$ SCFR/SFR.  Our derived $\Gamma$ values are listed in Table 3 for the representative combinations of IMF and metallicity, where we also report $\Gamma_{\rm corr}$, obtained by subtracting a 22.5\% of the derived value in order to account for possible field contamination, as detailed in \S\,3.6.

Reflecting the behaviour of our mass and age SSP fits, we find $\Gamma$ to be well constrained from below, irrespective of the assumptions of metallicity and IMF.  This is illustrated in the left-hand panel of Figure \ref{CFE_SFR}, which shows $\Gamma$ to be consistently around 30\% or higher for any combination of metallicity and IMF slope, although with large error bars.

In order to place the $\Gamma$ just derived for NGC 1427A within some context, we  will plot in the right-hand panel of Figure \ref{CFE_SFR}
 our measurements together with similar ones determined for spiral and starburst galaxies as a function of the global star formation rate (surface) density $\Sigma_{\rm SFR}$   \citep{Goddard:2010uq,Adamo:2011vn,Kruijssen:2012qf}.  These measurements seem to show that the cluster formation efficiency (SCFE) increases with $\Sigma_{\rm SFR}$, although with considerable scatter \citep{Cook:2012yq}, and there have even been reports of no clear trend at all \citep{Bastian:2008uq,Silva-Villa:2011rz}.

 {{}{A critical aspect during the derivation of $\Sigma_{\rm SFR}$ (x axis of the right panel of Fig \ref{CFE_SFR}) 
is the  proper selection of the area which it is defined. The choice of this area varies from author to author. For example, \cite{Adamo:2011vn} 
states that "To estimate a meaningful $\Sigma_{SFR}$ it is necessary to determine the size of the region which is producing stars"  and thus they assumed as normalising area  the  R$_{80\%}$. \cite{Cook:2012yq}  point out  that "The SFR surface density ($\Sigma_{SFR}$) is calculated  by normalising the SFR over the area of the galaxy being studied: Typically the whole galaxy in our sample."
Since we are resolving the stellar population in NGC 1427A, we have decided to compute its $\Sigma_{SFR}$ for two limiting cases given by the choice of the area involved in this quantity.  For the first one, %
we take into account the  R$_{80\%}$ of NGC 1427A as the area where the star formation occurs,
and thus we assume that the galaxy light is concentrated in this radius which corresponds to R=1.8 kpc as it would be the case of an unresolved blue galaxy.  However, complexes are located  across the
 "ring shape" of NGC 1427A  and therefore the star formation is  more localized in this area.  Hence, for the other extreme case we assume that all the star formation is occurring in the detected complexes, and therefore we have used the area covered by these complexes only (i.e. the effective radius) as the normalising area.
}

In the right-hand panel of Fig. \ref{CFE_SFR}, NGC 1427A is represented by filled black symbols with error bars (i.e., weighted means with the errors of those means). The corresponding values were obtained by simply
taking the error weighted average of the $\Gamma$ values for the 10 different [Fe/H]-IMF combinations shown in the left-hand panel.\textcolor{red}{\footnote{{}{y-axis error bars represent the uncertainties coming from the propagation of the SCFR and the SFR while for the x-axis the points are left without error since this corresponds to the two extreme cases.}}}  While such a procedure is strictly valid only for the statistical representation of a set of independent measurements  (i.e., unlike the case of our $\Gamma$ determinations, each of which represents the result of applying a model with different assumptions to the same set of observations), having the range of variation in $\Gamma$ available on the left-hand panel allows readers to make their own judgement about what value (or range) of $\Gamma$ best represents the situation.  The grey arrow in between the NGC 1427A data points represents the vertical shift applied as a correction for possible field contamination (\S\,3.6).   {{}{The filled triangle corresponds to the first limiting case where $\Sigma_{SFR}$ is calculated using the R$_{80\%}$ and the filled circle corresponds to the second limiting case where $\Sigma_{SFR}$ calculated considering the star complex effective radius.}}

{{}{When we consider both limiting cases we see that NGC 1427A shows a  large fraction of recent star formation occurring in star cluster and, depending on the aperture definition discussed above, it can be interpreted as a galaxy being significantly more efficient in forming star clusters (black filled triangle in Fig. \ref{CFE_SFR}), or in the alternative (black filled circle) case when only the small aperture is considered, it is comparable with star and star cluster formation rates found in blue compact starburst galaxies \citep{Adamo:2011vn}. We tend to prefer the latter interpretation because we are resolving the area where star formation actually occurs. However, we point out that the larger aperture choice would be more consistent for comparisons with star formation in distant, marginally resolved or unresolved high-z galaxies. In that case, our result hint at the tentative possibility that intermediate to high star-formation rates might be occurring in large fractions in star clusters in such systems.
} 
{{}{We think that regardless of the interpretation, the main result of the plots in Fig. 14 is the large $\Gamma$ derived here, along with (1) the peculiar location of this gas-rich galaxy so close in projection to the center of the Fornax cluster, and (2) the unusual morphology of these star forming regions being restricted to an edge of the galaxy, both very suggestive of a starburst triggered by the interaction between NGC 1427A and the Fornax galaxy cluster environment.
}

\begin{deluxetable}{ccccc}[!ht]
\tabletypesize{\scriptsize}
\tablecaption{SCFR and $\Gamma$ measured in NGC~1427A\label{TABLA_GAMMA}}
\tablewidth{0pt}
\tablehead{ 
\colhead{${\mathrm{IMF}}$ }&\colhead{[Fe/H]} &   \colhead{SCFR[M$_{\odot}$yr$^{-1}$]} &   \colhead{$\Gamma \%$} &  \colhead{$\Gamma\%_{corr} $} 
}
   		     &     -1.7   &      0.031$\pm$0.002    &        62$\pm$37             & 48 \\ 
   		     &     -0.7   &      0.027$\pm$0.003    &        53$\pm$33             & 41\\ 
SALPETER   &     -0.3   &      0.032$\pm$0.002    &        65$^{+35}_{-39}$   & 50\\  
		     &     0      &       0.034$\pm$0.002    &        69$^{+32}_ {-41}$  & 53\\ 
		     &     0.3    &      0.046$\pm$0.003    &        92$^{+8}_{-55}$     & 71\\ 
\hline
		   &     -1.7   &      0.021$\pm$0.003    &        42$\pm$26  &33\\ 
		   &     -0.7   &      0.017$\pm$0.002    &        35$\pm$21  &27\\ 
KROUPA     &     -0.3   &      0.020$\pm$0.001    &        41$\pm$25  &32\\ 
  		   &     0      &       0.022$\pm$0.001    &        43$\pm$26  &33\\ 
    		   &     0.3    &      0.029$\pm$0.002    &        59$\pm$35  &45

\enddata
\tablecomments{$\Gamma\%_{corr}=\Gamma-22.5$ as explained in section \S 3.7.}
\end{deluxetable}

\begin{figure*}
\centering
\includegraphics[width=16cm]{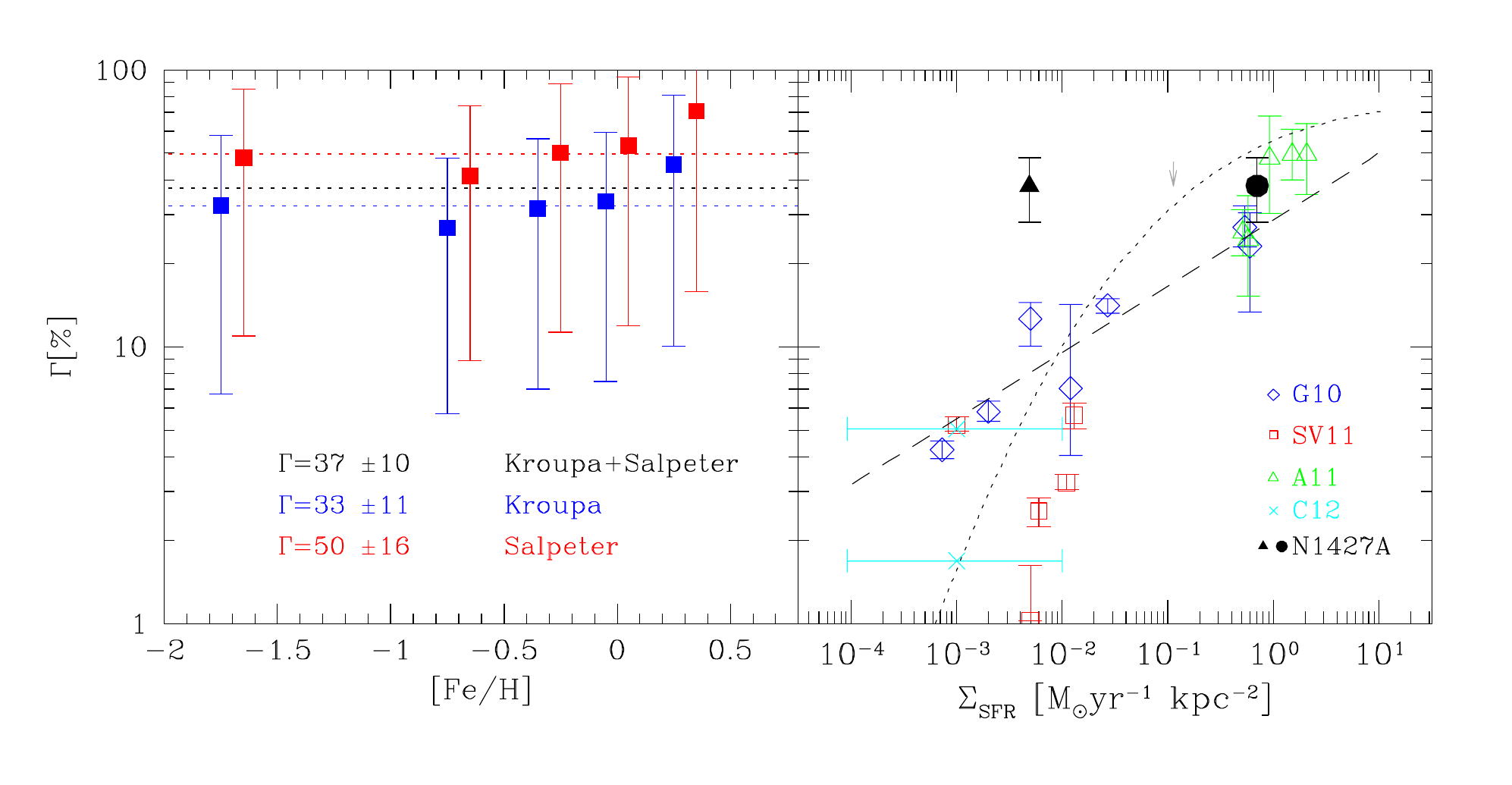}
\caption{Left: The variation in $\Gamma$ (corrected for contamination) at every combination of metallicity-IMF, including its corresponding uncertainty. 
For clarity, the corresponding points at each metallicity have been shifted by $\pm0.5$ dex around the real value ([Fe/H]=$0.3$, $0.0$, $-0.3$, $-0.7$ and $-1.7$ dex). Blue filled squares correspond to Kroupa IMF and red filled squares correspond to Salpeter IMF calculations. Horizontal dotted lines represent weighted mean for $\Gamma$s considering Kroupa IMF (blue), Salpeter IMF (red) and both IMF types (black). Right: $\Gamma$ against Star Formation Rate (SFR) surface density. Open blue diamonds correspond to the samples from \cite{Goddard:2010uq}, green triangles to \cite{Adamo:2011vn}, red squares to \cite{Silva-Villa:2011rz}. The cyan crosses correspond to 
the sample of \cite{Cook:2012yq}. The doted curve represents the modelled relation  for the "typical" parameters  set from the table $1$ of \cite{Kruijssen:2012qf} and the dashed line represents the fit from \cite{Goddard:2010uq}. The filled black symbols correspond to the weighted mean (corrected for contamination) for NGC 1427A considering both IMFs with different normalising areas: R$_{80\%}$ represented as a black triangle and, the star complex effective radius,  represented as a black circle (for further details see text). The vertical error bars correspond to the error of the mean for $\Gamma$. 
The grey arrow in between the filled black dots represents the $22.5\%$  displacement of correction due to contamination of $\Gamma$ (see \S. 3.6)}
\label{CFE_SFR}
\end{figure*}

\section {Discussion and conclusions}

We have analyzed deep multicolor HST observations of
NGC 1427A, a Magellanic-like, gas-rich dIrr galaxy near the core of
the Fornax cluster, and characterized the properties of its most
recent episode of star formation.  At the distance of the Fornax
cluster, this means studying the population of resolved young star
clusters and star-cluster complexes, as well as the unresolved,
diffuse young field stellar population traced by H$\alpha$ emission.

{A set of 12 sources stand out clearly due to their
extreme blue colors, and we find that they correspond to relatively
extended structures similar to star-cluster complexes already seen in
other galaxies.  ~The signature of hierarchical formation processes is
evident in several of the star cluster complexes, where smaller star
clusters are members of larger star-forming complexes.  Almost all
star cluster complexes seem to be associated with other young
entities, as illustrated in the postage-stamp images in
Figure~\ref{sample} which show numerous nearby blue (i.e.~young)
objects. This is consistent with the fractal clustering of stars, star
clusters and star cluster complexes seen in other dwarf galaxies
(e.g.~LMC: \citealt{Bastian:2009kx}; NGC\,6822:
\citealt{Gouliermis:2010vn}) and numerical simulations
\citep[e.g.][and references there in]{Allison:2010ys}.}

\begin{figure}
\centering
\includegraphics[width=8.5cm]{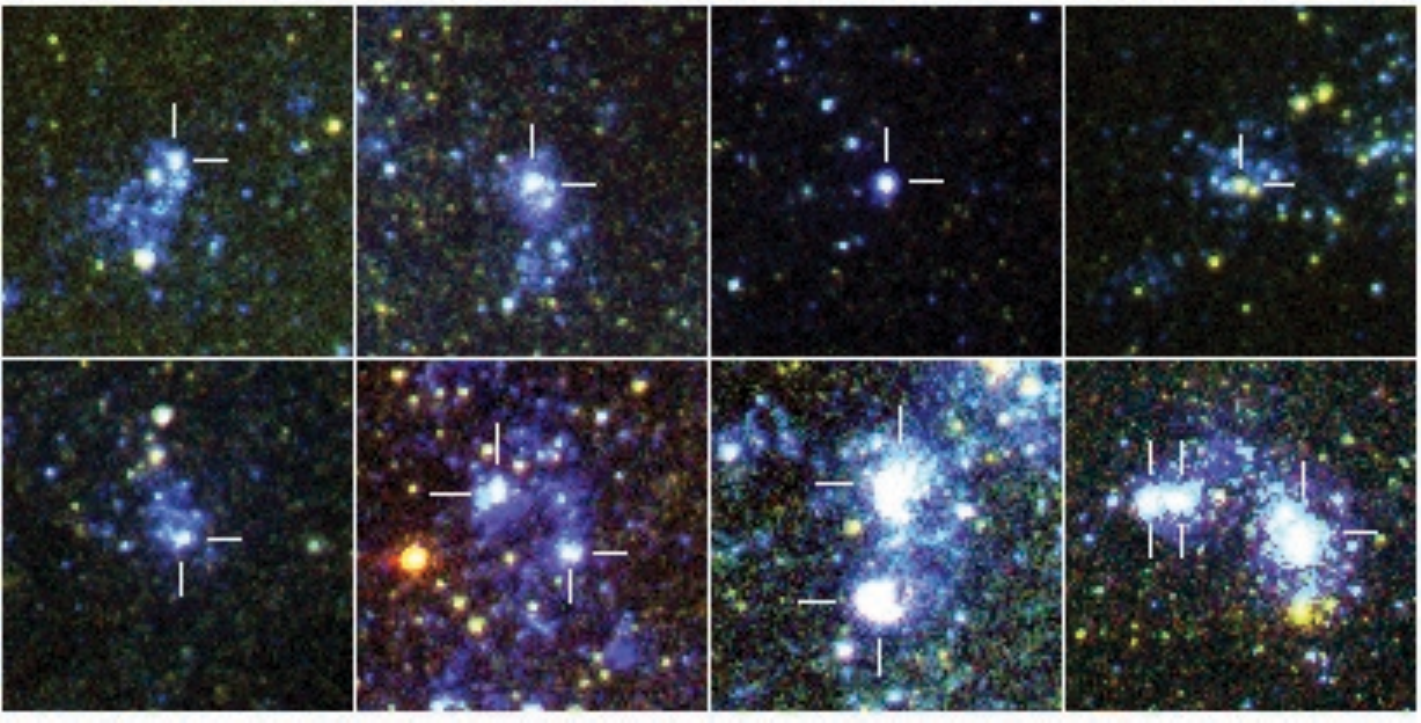}
\caption{HST/ACS color composite mosaic of the star cluster
complexes. The color code used for blue corresponds to the $F475W$,
for green corresponds to $F775W$ and for red corresponds to the
$F850LP$ passband. Each stamp represents a field of view of $402
\times$402 pc$^{2}$. Each star cluster complex is indicated by two
white lines. From left to right in Top row: ID:891, ID:1163, ID: 5872,
and ID:6003. Bottom row left stamp: ID3364. Each one of the two bottom
central panels shows 2 star cluster complexes: ID:2892 and ID:3000
(left and right) and ID:1383 and ID:1776. The panel at bottom right
shows 3 star cluster complexes (from left to right): ID:3070, ID:3090,
and ID:3464.}
\label{sample}
\end{figure}

~{The sizes of the} star cluster complexes
are somewhat filter dependent, probably because these structures are
composed of a mix of several ingredients: a mixture of gas, OB
associations, and star cluster stellar populations, the latter of
which are likely to be significantly affected by stochastic
fluctuations in their stellar content.  \cite{Fouesneau:2010uq}.
 studied these stochastic fluctuations on the recovered cluster mass, and these show a scatter around the one-to-one line that is slightly smaller than $\sim$ 0.5 dex at the mass range of the clusters we study ($\sim 10^{3.5}$ M$_\odot$; see their Figure 3).  The fact, however, that this scatter is symmetric throughout the range of cluster mass relevant for our study means that these particular stochasticity effects will more or less cancel out when the cluster masses are added up for the computation of $\Gamma$ (i.e., about half of our clusters will have their mass overestimated and half will have it underestimated).  {{Further modelling of star cluster complexes and its dependence on the sampling of their mass function
are beyond the limits of the data  since we are  affected by sampling effects and  therefore we have no access to global information of 
the complexes  but only particular information which is  dominated by few individual objects \citep{Villaverde:2009pd}.}}

Seven of the detected star cluster complexes are observed relatively
close in projection and two of them (IDs:~3070 and ID:~3090, see
Figure~\ref{sample}) are the pair that appear closest to one another
with a projected distance $\sim\!28$ pc that can be considered as the
NGC\,1427A counterparts of star cluster pairs
{(possibly bound binary systems)} seen in the LMC
\citep[e.g.][]{Dieball:2002zr, Dieball:2000ly}. However, due to their
location and the nature of the host galaxy, it is likely that they
will be disrupted in the near future either by cluster-cluster
interactions, ISM interactions \citep{Gieles:2006fk} or/and the
interactions with the Fornax galaxy cluster.

{Based on the H$\alpha$ flux measured over the entire
galaxy, and accounting for the possibility of undetected H$\alpha$
flux, we compute the global SFR in NGC 1427A and find it similar to
the mean of star-forming galaxies of similar type}, such as those
studied by \cite{Seth:2004}. {Then, using our
determinations of the masses and ages of the star cluster complexes,
we compute the SCFR, i.e., the amount of recent star formation
occurring within star clusters.  {{}{We find the ratio $\Gamma$ of these
two SFRs to be  large and consistent with observations of 
starburst galaxies  \citep[e.g.][]{Adamo:2011vn}.  Therefore, the star cluster complexes in NGC 1427A indicate that there is an ongoing star formation episode during which a large fraction of the star formation over the last few million years occurred in star clusters.}}

{{}{It would be extremely interesting to determine
whether or not the recent  star cluster formation
in NGC 1427A has been triggered by interaction with the cluster
environment.  This could be achieved by a measurement of $\Gamma$, the
SCFR relative to the field SFR, for star formation episodes that
occurred before the galaxy started its plunge into the dense core of
the Fornax cluster, i.e., for stellar populations with ages comparable
and larger than a crossing time through the Fornax center.  This is,
however, a complex task for which not only the star and
cluster formation histories would have to be derived, but also the
star cluster destruction processes (e.g., infant mortality) would need
to be accounted for.}}

\acknowledgments
\noindent {\it Acknowledgements} -- {We gratefully acknowledge the thoughtful comments and suggestions of the anonymous referee. {{} { We also acknowledge Angela Adamo, Diederik Kruijssen and Nate Bastian for insightful discussions regarding the various star formation density measurement approaches.}}
~This research was supported by CONICYT through FONDECYT/Regular Project No.~1121005, FONDAP Center for Astrophysics (15010003)  FONDECYT/Regular Project  No.~1130373 and by BASAL PFB-06 "Centro de Astronom\'ia y Tecnolog\'ias Afines''; and by the Ministry for the Economy, Development, and Tourism's Programa Iniciativa Cient\'ifica Milenio through grant IC 120009, awarded to the Millennium Institute of Astrophysics (MAS).~M.~D.~Mora is grateful for support from Gemini-CONICYT.

{\it Facilities:} \facility{HST (ACS)}, \facility{ESO/VLT (FORS1)}.



\appendix

{
\section{Appendix material}

\tiny
\scriptsize
\clearpage

\begin{table}
\caption{Star cluster complex parameters I: Sizes and Magnitudes}
\centering
\begin{tabular}{c|ccccccccc}
\hline\hline
ID &   R$_\mathrm{eff}$      &  R$_\mathrm{eff}$   & R$_\mathrm{eff}$     & R$_\mathrm{eff}$     &  FORS & ACS & ACS & ACS & ACS \\
     & F475W   & F625W& F775W & F850LP & U  & F475W &    F625W  &  F775W  &  F850LP  \\
\hline\hline
 891   & 63.04  &  72.60  & 526.8  & 346.3   &   23.412$\pm$0.459  &  22.177$\pm$0.006  &  22.131$\pm$0.008  &  22.746$\pm$0.023  & 22.447$\pm$0.030    \\ 
 1163 & 14.38  &  78.06  & 6.46    & 11.74    &   20.726$\pm$0.319  &  21.962$\pm$0.006  &  20.829$\pm$0.003  &  22.562$\pm$0.020  & 22.254$\pm$0.026    \\ 
 1383 & 54.11  &  87.28  & 11.91   &  73.25   &   20.578$\pm$0.188  &  20.952$\pm$0.003  &  20.541$\pm$0.003  &  22.124$\pm$0.016  & 20.908$\pm$0.009    \\
 1776 & 24.59  &  28.94  & 17.61   &  28.11   &   20.862$\pm$0.125  &  20.693$\pm$0.002  &  20.653$\pm$0.003  &  21.614$\pm$0.009  & 20.968$\pm$0.009    \\
 2892 & 18.37  &  24.36  & 20.89   &  15.96   &   20.703$\pm$0.054  &  21.875$\pm$0.005  &  21.484$\pm$0.005  &  22.206$\pm$0.015  & 22.206$\pm$0.026    \\
 3000 & 14.03  &  19.54  & 6.63    &  8.39    &   21.182$\pm$0.031  &  22.419$\pm$0.007  &  21.938$\pm$0.008  &  22.910$\pm$0.027  & 22.720$\pm$0.040    \\
 3070 & 6.57    &   5.63   & 0.70    &  0.12    &   20.568$\pm$0.031  &  22.001$\pm$0.006  &  22.186$\pm$0.010  &  22.556$\pm$0.020  & 22.532$\pm$0.033    \\
 3090 & 17.14  &  14.67  & 16.73   &  16.61   &   21.488$\pm$0.048  &  22.118$\pm$0.007  &  22.026$\pm$0.008  &  22.545$\pm$0.019  & 22.246$\pm$0.025    \\
 3364 & 12.09  &  11.68  & 11.39   &  8.80    &   20.966$\pm$0.042  &  22.235$\pm$0.007  &  22.305$\pm$0.012  &  22.516$\pm$0.021  & 22.679$\pm$0.041    \\
 3464 & 17.78  &  17.61  & 11.09   &  8.39    &   20.287$\pm$0.039  &  21.481$\pm$0.004  &  21.404$\pm$0.006  &  21.727$\pm$0.010  & 21.709$\pm$0.017    \\
 5872 & 3.82    &   3.82   & 2.88    &  4.05    &   21.555$\pm$0.034  &  22.275$\pm$0.006  &  22.245$\pm$0.009  &  23.069$\pm$0.026  & 22.681$\pm$0.033    \\
 6002 & 147.0  &  14.61  & 4.40    &  0.12    &   22.517$\pm$0.269  &  21.467$\pm$0.003  &  22.890$\pm$0.017  &  22.577$\pm$0.020  & 22.283$\pm$0.027    \\
\hline
\end{tabular}
\tablecomments{Star cluster sizes are given in units of parsec. Luminosities are given in units of magnitudes. For a King30 profile R$_\mathrm{eff} = 1.48\times$FWHM}
\label{tablefinal}
\end{table}

\footnotesize

\begin{table}
\footnotesize
\caption{Star cluster complexes parameters II: Extinctions, Ages, and Masses for Kroupa and Salpeter IMF}
\centering
\rotatebox{90}{
\begin{tabular}{c|ccc|ccc|ccc|ccc|ccc}
\hline\hline
   &\multicolumn{3}{c|}{[Fe/H]~$=-1.7$}     & \multicolumn{3}{c|}{[Fe/H]~$=-0.7$}  
   &\multicolumn{3}{c|}{[Fe/H]~$=-0.3$}     & \multicolumn{3}{c|}{[Fe/H]~$=0.0$}   
   &\multicolumn{3}{c}{[Fe/H]~$=+0.3$}       \\
ID &  E(B-V)$^{\mathrm{max}}_{\mathrm{min}}$   & Age$^{\mathrm{max}}_{\mathrm{min}}$    & Mass$^{\mathrm{max}}_{\mathrm{min}}$ &  E(B-V)$^{\mathrm{max}}_{\mathrm{min}}$   & Age$^{\mathrm{max}}_{\mathrm{min}}$    & Mass$^{\mathrm{max}}_{\mathrm{min}}$ &  E(B-V)$^{\mathrm{max}}_{\mathrm{min}}$     & Age$^{\mathrm{max}}_{\mathrm{min}}$    & Mass$^{\mathrm{max}}_{\mathrm{min}}$ &  E(B-V)$^{\mathrm{max}}_{\mathrm{min}}$     & Age$^{\mathrm{max}}_{\mathrm{min}}$    & Mass$^{\mathrm{max}}_{\mathrm{min}}$ &  E(B-V)$^{\mathrm{max}}_{\mathrm{min}}$     & Age$^{\mathrm{max}}_{\mathrm{min}}$    & Mass$^{\mathrm{max}}_{\mathrm{min}}$ \\
\hline\hline
\multicolumn{16}{c}{Kroupa IMF}\\\hline
891  & 0.00$^{0.02}_{0.00}$ & 6.60$^{0.01}_{0.00}$ & 3.31$^{0.02}_{0.00}$ & 0.10$^{0.04}_{0.06}$ & 6.60$^{0.02}_{0.00}$ & 3.60$^{0.06}_{0.07}$ & 0.00$^{0.04}_{0.00}$ & 6.60$^{0.03}_{0.00}$ & 3.63$^{0.05}_{0.00}$ & 0.00$^{0.04}_{0.00}$ & 6.60$^{0.03}_{0.00}$ & 3.67$^{0.04}_{0.00}$ & 0.00$^{0.04}_{0.00}$ & 6.60$^{0.01}_{0.00}$ & 3.77$^{0.04}_{0.00}$ \\
1163 & 0.04$^{0.04}_{0.04}$ & 6.90$^{0.01}_{0.30}$ & 4.35$^{0.05}_{0.83}$ & 0.00$^{0.12}_{0.02}$ & 6.60$^{0.01}_{0.00}$ & 3.87$^{0.07}_{0.05}$ & 0.00$^{0.00}_{0.00}$ & 6.60$^{0.01}_{0.00}$ & 3.84$^{0.00}_{0.00}$ & 0.00$^{0.00}_{0.00}$ & 6.60$^{0.01}_{0.00}$ & 3.87$^{0.01}_{0.00}$ & 0.00$^{0.00}_{0.00}$ & 6.60$^{0.00}_{0.00}$ & 3.97$^{0.01}_{0.00}$ \\
1383 & 0.00$^{0.02}_{0.00}$ & 6.60$^{0.00}_{0.00}$ & 3.82$^{0.02}_{0.00}$ & 0.08$^{0.06}_{0.04}$ & 6.60$^{0.00}_{0.00}$ & 4.09$^{0.07}_{0.05}$ & 0.00$^{0.02}_{0.00}$ & 6.60$^{0.01}_{0.00}$ & 4.14$^{0.02}_{0.00}$ & 0.00$^{0.02}_{0.00}$ & 6.61$^{0.03}_{0.01}$ & 4.17$^{0.03}_{0.00}$ & 0.00$^{0.04}_{0.00}$ & 6.60$^{0.01}_{0.00}$ & 4.27$^{0.05}_{0.00}$ \\
1776 & 0.00$^{0.02}_{0.00}$ & 6.60$^{0.00}_{0.00}$ & 3.86$^{0.03}_{0.00}$ & 0.10$^{0.04}_{0.04}$ & 6.60$^{0.01}_{0.00}$ & 4.15$^{0.05}_{0.04}$ & 0.00$^{0.06}_{0.00}$ & 6.60$^{0.02}_{0.00}$ & 4.18$^{0.07}_{0.00}$ & 0.00$^{0.04}_{0.00}$ & 6.62$^{0.03}_{0.02}$ & 4.22$^{0.04}_{0.01}$ & 0.02$^{0.04}_{0.02}$ & 6.60$^{0.01}_{0.00}$ & 4.33$^{0.05}_{0.02}$ \\
2892 & 0.06$^{0.04}_{0.04}$ & 6.90$^{0.01}_{0.00}$ & 4.35$^{0.05}_{0.05}$ & 0.00$^{0.04}_{0.00}$ & 6.61$^{0.02}_{0.01}$ & 3.68$^{0.05}_{0.01}$ & 0.00$^{0.02}_{0.00}$ & 6.60$^{0.02}_{0.00}$ & 3.81$^{0.03}_{0.00}$ & 0.00$^{0.02}_{0.00}$ & 6.60$^{0.02}_{0.00}$ & 3.85$^{0.02}_{0.00}$ & 0.00$^{0.04}_{0.00}$ & 6.60$^{0.01}_{0.00}$ & 3.94$^{0.05}_{0.00}$ \\
3000 & 0.02$^{0.04}_{0.02}$ & 6.90$^{0.01}_{0.00}$ & 4.09$^{0.05}_{0.03}$ & 0.00$^{0.00}_{0.00}$ & 6.60$^{0.01}_{0.00}$ & 3.46$^{0.01}_{0.00}$ & 0.00$^{0.00}_{0.00}$ & 6.60$^{0.01}_{0.00}$ & 3.60$^{0.01}_{0.00}$ & 0.00$^{0.02}_{0.00}$ & 6.60$^{0.01}_{0.00}$ & 3.63$^{0.03}_{0.00}$ & 0.00$^{0.02}_{0.00}$ & 6.60$^{0.01}_{0.00}$ & 3.73$^{0.03}_{0.00}$ \\
3070 & 0.00$^{0.00}_{0.00}$ & 6.90$^{0.02}_{0.00}$ & 4.16$^{0.02}_{0.00}$ & 0.00$^{0.00}_{0.00}$ & 6.60$^{0.01}_{0.00}$ & 3.56$^{0.01}_{0.00}$ & 0.00$^{0.00}_{0.00}$ & 6.60$^{0.01}_{0.00}$ & 3.70$^{0.00}_{0.00}$ & 0.00$^{0.00}_{0.00}$ & 6.60$^{0.00}_{0.00}$ & 3.73$^{0.00}_{0.00}$ & 0.00$^{0.00}_{0.00}$ & 6.60$^{0.00}_{0.00}$ & 3.82$^{0.01}_{0.00}$ \\
3090 & 0.00$^{0.02}_{0.00}$ & 6.60$^{0.01}_{0.00}$ & 3.37$^{0.03}_{0.00}$ & 0.14$^{0.04}_{0.04}$ & 6.60$^{0.03}_{0.00}$ & 3.71$^{0.05}_{0.05}$ & 0.06$^{0.08}_{0.06}$ & 6.64$^{0.07}_{0.04}$ & 3.77$^{0.08}_{0.06}$ & 0.00$^{0.14}_{0.00}$ & 6.67$^{0.03}_{0.07}$ & 3.73$^{0.15}_{0.01}$ & 0.14$^{0.04}_{0.04}$ & 6.60$^{0.01}_{0.00}$ & 3.98$^{0.05}_{0.06}$ \\
3364 & 0.00$^{0.00}_{0.00}$ & 7.08$^{0.01}_{0.14}$ & 4.31$^{0.00}_{0.19}$ & 0.00$^{0.00}_{0.00}$ & 6.65$^{0.03}_{0.03}$ & 3.55$^{0.03}_{0.04}$ & 0.00$^{0.00}_{0.00}$ & 6.60$^{0.03}_{0.00}$ & 3.63$^{0.03}_{0.00}$ & 0.00$^{0.00}_{0.00}$ & 6.60$^{0.01}_{0.00}$ & 3.66$^{0.01}_{0.00}$ & 0.00$^{0.00}_{0.00}$ & 6.60$^{0.01}_{0.00}$ & 3.76$^{0.01}_{0.00}$ \\
3464 & 0.00$^{0.02}_{0.00}$ & 7.08$^{0.02}_{0.00}$ & 4.63$^{0.04}_{0.00}$ & 0.00$^{0.00}_{0.00}$ & 6.66$^{0.03}_{0.02}$ & 3.88$^{0.03}_{0.02}$ & 0.00$^{0.02}_{0.00}$ & 6.63$^{0.03}_{0.03}$ & 3.98$^{0.02}_{0.02}$ & 0.00$^{0.04}_{0.00}$ & 6.61$^{0.02}_{0.01}$ & 3.99$^{0.05}_{0.00}$ & 0.02$^{0.04}_{0.02}$ & 6.60$^{0.02}_{0.00}$ & 4.11$^{0.05}_{0.02}$ \\
5872 & 0.08$^{0.02}_{0.04}$ & 6.90$^{0.01}_{0.00}$ & 4.12$^{0.03}_{0.06}$ & 0.00$^{0.04}_{0.00}$ & 6.60$^{0.01}_{0.00}$ & 3.41$^{0.06}_{0.00}$ & 0.00$^{0.02}_{0.00}$ & 6.60$^{0.02}_{0.00}$ & 3.55$^{0.03}_{0.00}$ & 0.00$^{0.04}_{0.00}$ & 6.61$^{0.02}_{0.01}$ & 3.58$^{0.06}_{0.00}$ & 0.02$^{0.04}_{0.02}$ & 6.60$^{0.01}_{0.00}$ & 3.70$^{0.06}_{0.02}$ \\
6002 & 0.00$^{0.00}_{0.00}$ & 7.08$^{0.01}_{0.00}$ & 4.34$^{0.01}_{0.00}$ & 0.00$^{0.00}_{0.00}$ & 6.65$^{0.04}_{0.02}$ & 3.57$^{0.04}_{0.02}$ & 0.00$^{0.00}_{0.00}$ & 6.62$^{0.02}_{0.02}$ & 3.68$^{0.01}_{0.01}$ & 0.00$^{0.00}_{0.00}$ & 6.60$^{0.01}_{0.00}$ & 3.70$^{0.00}_{0.00}$ & 0.00$^{0.00}_{0.00}$ & 6.60$^{0.02}_{0.00}$ & 3.80$^{0.01}_{0.00}$ \vspace{0.1cm}\\\hline
\multicolumn{16}{c}{Salpeter IMF}\\\hline
891  & 0.00$^{0.02}_{0.00}$ & 6.60$^{0.01}_{0.00}$ & 3.52$^{0.02}_{0.00}$ & 0.10$^{0.04}_{0.06}$ & 6.60$^{0.02}_{0.00}$ & 3.80$^{0.06}_{0.07}$ & 0.00$^{0.04}_{0.00}$ & 6.60$^{0.03}_{0.00}$ & 3.83$^{0.05}_{0.00}$ & 0.00$^{0.04}_{0.00}$ & 6.60$^{0.03}_{0.00}$ & 3.87$^{0.04}_{0.00}$ & 0.00$^{0.04}_{0.00}$ & 6.60$^{0.01}_{0.00}$ & 3.96$^{0.05}_{0.00}$ \\
1163 & 0.00$^{0.00}_{0.00}$ & 6.60$^{0.00}_{0.00}$ & 3.72$^{0.01}_{0.00}$ & 0.00$^{0.06}_{0.00}$ & 6.60$^{0.01}_{0.00}$ & 3.89$^{0.07}_{0.00}$ & 0.00$^{0.00}_{0.00}$ & 6.60$^{0.01}_{0.00}$ & 4.04$^{0.00}_{0.00}$ & 0.00$^{0.00}_{0.00}$ & 6.60$^{0.01}_{0.00}$ & 4.08$^{0.00}_{0.00}$ & 0.00$^{0.00}_{0.00}$ & 6.60$^{0.00}_{0.00}$ & 4.17$^{0.00}_{0.00}$ \\
1383 & 0.00$^{0.02}_{0.00}$ & 6.60$^{0.00}_{0.00}$ & 4.03$^{0.02}_{0.00}$ & 0.08$^{0.06}_{0.04}$ & 6.60$^{0.00}_{0.00}$ & 4.29$^{0.07}_{0.05}$ & 0.00$^{0.02}_{0.00}$ & 6.60$^{0.01}_{0.00}$ & 4.34$^{0.02}_{0.00}$ & 0.00$^{0.02}_{0.00}$ & 6.62$^{0.03}_{0.02}$ & 4.38$^{0.02}_{0.01}$ & 0.00$^{0.04}_{0.00}$ & 6.60$^{0.01}_{0.00}$ & 4.47$^{0.04}_{0.00}$ \\
1776 & 0.00$^{0.02}_{0.00}$ & 6.60$^{0.00}_{0.00}$ & 4.07$^{0.02}_{0.00}$ & 0.10$^{0.04}_{0.04}$ & 6.60$^{0.01}_{0.00}$ & 4.36$^{0.05}_{0.05}$ & 0.00$^{0.06}_{0.00}$ & 6.60$^{0.02}_{0.00}$ & 4.38$^{0.07}_{0.00}$ & 0.00$^{0.04}_{0.00}$ & 6.63$^{0.02}_{0.03}$ & 4.42$^{0.04}_{0.01}$ & 0.02$^{0.04}_{0.02}$ & 6.60$^{0.01}_{0.00}$ & 4.53$^{0.05}_{0.02}$ \\
2892 & 0.06$^{0.04}_{0.04}$ & 6.90$^{0.01}_{0.00}$ & 4.54$^{0.05}_{0.05}$ & 0.00$^{0.04}_{0.00}$ & 6.61$^{0.02}_{0.01}$ & 3.88$^{0.06}_{0.00}$ & 0.00$^{0.02}_{0.00}$ & 6.60$^{0.02}_{0.00}$ & 4.02$^{0.02}_{0.00}$ & 0.00$^{0.02}_{0.00}$ & 6.60$^{0.02}_{0.00}$ & 4.05$^{0.02}_{0.00}$ & 0.00$^{0.02}_{0.00}$ & 6.60$^{0.01}_{0.00}$ & 4.14$^{0.02}_{0.00}$ \\
3000 & 0.02$^{0.04}_{0.02}$ & 6.90$^{0.01}_{0.00}$ & 4.27$^{0.06}_{0.02}$ & 0.00$^{0.00}_{0.00}$ & 6.60$^{0.01}_{0.00}$ & 3.67$^{0.01}_{0.00}$ & 0.00$^{0.00}_{0.00}$ & 6.60$^{0.01}_{0.00}$ & 3.80$^{0.01}_{0.00}$ & 0.00$^{0.02}_{0.00}$ & 6.60$^{0.01}_{0.00}$ & 3.83$^{0.03}_{0.00}$ & 0.00$^{0.02}_{0.00}$ & 6.60$^{0.01}_{0.00}$ & 3.92$^{0.03}_{0.00}$ \\
3070 & 0.00$^{0.00}_{0.00}$ & 6.90$^{0.03}_{0.00}$ & 4.34$^{0.03}_{0.00}$ & 0.00$^{0.00}_{0.00}$ & 6.60$^{0.01}_{0.00}$ & 3.76$^{0.01}_{0.00}$ & 0.00$^{0.00}_{0.00}$ & 6.60$^{0.01}_{0.00}$ & 3.90$^{0.00}_{0.00}$ & 0.00$^{0.00}_{0.00}$ & 6.60$^{0.00}_{0.00}$ & 3.93$^{0.00}_{0.00}$ & 0.00$^{0.00}_{0.00}$ & 6.60$^{0.01}_{0.00}$ & 4.02$^{0.00}_{0.00}$ \\
3090 & 0.00$^{0.02}_{0.00}$ & 6.60$^{0.01}_{0.00}$ & 3.58$^{0.03}_{0.00}$ & 0.14$^{0.04}_{0.04}$ & 6.60$^{0.04}_{0.00}$ & 3.91$^{0.06}_{0.05}$ & 0.06$^{0.08}_{0.06}$ & 6.64$^{0.08}_{0.04}$ & 3.97$^{0.08}_{0.06}$ & 0.00$^{0.14}_{0.00}$ & 6.68$^{0.02}_{0.08}$ & 3.93$^{0.15}_{0.01}$ & 0.14$^{0.04}_{0.04}$ & 6.60$^{0.01}_{0.00}$ & 4.17$^{0.05}_{0.05}$ \\
3364 & 0.00$^{0.00}_{0.00}$ & 7.08$^{0.01}_{0.15}$ & 4.49$^{0.00}_{0.19}$ & 0.00$^{0.00}_{0.00}$ & 6.65$^{0.03}_{0.02}$ & 3.75$^{0.03}_{0.03}$ & 0.00$^{0.00}_{0.00}$ & 6.60$^{0.03}_{0.00}$ & 3.83$^{0.03}_{0.00}$ & 0.00$^{0.00}_{0.00}$ & 6.60$^{0.01}_{0.00}$ & 3.86$^{0.01}_{0.00}$ & 0.00$^{0.00}_{0.00}$ & 6.60$^{0.01}_{0.00}$ & 3.95$^{0.01}_{0.00}$ \\
3464 & 0.00$^{0.02}_{0.00}$ & 7.08$^{0.02}_{0.00}$ & 4.81$^{0.04}_{0.00}$ & 0.00$^{0.00}_{0.00}$ & 6.67$^{0.02}_{0.03}$ & 4.09$^{0.03}_{0.03}$ & 0.00$^{0.02}_{0.00}$ & 6.63$^{0.03}_{0.03}$ & 4.18$^{0.02}_{0.02}$ & 0.00$^{0.04}_{0.00}$ & 6.60$^{0.03}_{0.00}$ & 4.19$^{0.05}_{0.00}$ & 0.02$^{0.04}_{0.02}$ & 6.60$^{0.02}_{0.00}$ & 4.31$^{0.05}_{0.03}$ \\
5872 & 0.08$^{0.02}_{0.04}$ & 6.90$^{0.01}_{0.00}$ & 4.30$^{0.03}_{0.05}$ & 0.00$^{0.04}_{0.00}$ & 6.60$^{0.01}_{0.00}$ & 3.61$^{0.06}_{0.00}$ & 0.00$^{0.02}_{0.00}$ & 6.60$^{0.02}_{0.00}$ & 3.75$^{0.03}_{0.00}$ & 0.00$^{0.04}_{0.00}$ & 6.61$^{0.02}_{0.01}$ & 3.79$^{0.05}_{0.01}$ & 0.02$^{0.04}_{0.02}$ & 6.60$^{0.01}_{0.00}$ & 3.90$^{0.05}_{0.03}$ \\
6002 & 0.00$^{0.00}_{0.00}$ & 7.08$^{0.01}_{0.00}$ & 4.52$^{0.01}_{0.00}$ & 0.00$^{0.00}_{0.00}$ & 6.65$^{0.04}_{0.02}$ & 3.77$^{0.04}_{0.02}$ & 0.00$^{0.00}_{0.00}$ & 6.61$^{0.03}_{0.01}$ & 3.87$^{0.02}_{0.00}$ & 0.00$^{0.00}_{0.00}$ & 6.60$^{0.01}_{0.00}$ & 3.90$^{0.00}_{0.00}$ & 0.00$^{0.00}_{0.00}$ & 6.60$^{0.02}_{0.00}$ & 3.99$^{0.02}_{0.00}$ \\
\hline
\end{tabular}
}
\tablecomments{The total extinction values are given in units of magnitudes, while star cluster ages and masses are provided in units of the decadal logarithm of years and solar masses, respectively.}
\label{tablefinal2}

\end{table}


\normalsize



}
\bibliographystyle{aa}

\bibliography{Marcelo}

\end{document}